\documentclass[proof]{WileyASNA-v1}

\usepackage{color}

\articletype{Article Type}%

\received{August 25, 2022}
\revised{January 12, 2023}
\accepted{January 12, 2023}

\begin{document}

\title{Surface magnetic field of the A-type metallic-line star omicron~Pegasi revisited}

\author[]{Yoichi Takeda}

\authormark{Y. TAKEDA}

\address[]{ 
\orgaddress{\state{11-2 Enomachi, Naka-ku, Hiroshima-shi, 730-0851}, \country{Japan}}}

\corres{\email{ytakeda@js2.so-net.ne.jp}}

%\presentaddress{This is sample for present address text this is sample for present address text}

\abstract{
The bright A-type metallic-line star $o$~Peg was reported in the early 1990s
to have a surface magnetic field of $\sim 2$~kG 
by analyzing the widths and strengths of spectral lines. 
In respect that those old studies were of rather empirical 
or approximate nature and the quality of observational data was not sufficient, 
this problem has been newly reinvestigated based on physically 
more rigorous simulations of line flux profiles, along with the observed equivalent 
widths ($W$) and full-widths at half-maximum ($h$) of 198 Fe~{\sc i} and 
182 Fe~{\sc ii} lines measured from the high-quality spectra. 
Given the Fe abundance derived from the conventional analysis, 
theoretical $W$ and $h$ values calculated for various sets of parameters were 
compared with the observed ones, which lead to the following conclusion 
regarding $\langle H \rangle$ (mean field strength).
(1) An analysis of $W$ yielded $\langle H \rangle \sim$1--1.5~kG from Fe~{\sc ii} lines 
with the microturbulence of $v_{\rm t} \sim 1.5$~km~s$^{-1}$.
(2) A comparison of $h$ resulted in $\langle H \rangle \sim$1.5--2~kG as well as 
the projected rotational velocity of $v_{\rm e}\sin i \simeq 5$~km~s$^{-1}$.
(3)  Accordingly, the existence of mean magnetic field on the order of
$\langle H \rangle \sim$~1--2~kG in $o$~Peg was confirmed, which is almost 
consistent with the consequence of the previous work. 
}


\maketitle

\footnotetext{\textbf{Abbreviations:} LTE, local thermodynamic equilibrium; 
FWHM, full width at half maximum}

%Sect. 1
\section{Introduction}

The star $o$ Peg (= HD~214994 = HR~8641 = HIP~112051; spectral type is A1~IV) 
is one of the most frequently studied A-type stars in stellar spectroscopy, 
because of its apparent brightness ($V = 4.79$) and sharp-line nature with 
low projected rotational velocity ($v_{\rm e}\sin i \lesssim 10$~km~s$^{-1}$).
As is often the case with slowly-rotating stars in the upper main sequence, it 
is a chemically peculiar (CP) star; more specifically, because of the 
conspicuous overabundances in the elements heavier than the Fe group, 
it is regarded as a hot metallic-lined A-type (Am) star (e.g., Adelman 1988, 
Hill \& Landstreet 1993). Definite variability has not been confirmed,
despite of the designation ``Si-variable'' and ``variable radial velocity'' 
in Renson \& Manfroid's (2009) catalogue of CP stars.   

A remarkable and rare feature of this star is that a surface magnetic field 
was spectroscopically detected in the past, despite that Am stars are 
generally not magnetic (unlike other groups of CP stars such as the SrCrEu type).
\footnote{
The meaning of this statement (non-magnetic nature of Am stars in general)
is that appreciable magnetism (e.g., field strength on the order of $\sim$kG) 
found in magnetic CP stars is generally absent in Am stars. Recent very high-precision 
spectropolarimetric observations have revealed that an extremely weak magnetic field
(on the order of $\sim$G) is detectable in several hot Am stars; 
Sirius (Petit et al. 2009), $\beta$~UMa and $\theta$~Leo (Blaz\`{e}re et al. 2016a),
Alhena ($\gamma$~Gem) (Blaz\`{e}re et al. 2016b, 2020).
}
\\  
--- It was Mathys \& Lanz (1990; hereinafter referred to as ML90) who first
reported the existence of a magnetic field of $H \sim 2$~kG in $o$~Peg,
where they employed two independent techniques: (i) statistical line-width 
analysis (Stenflo \& Lindegren 1977) and (ii) empirical relation for 
the $H$-dependent difference of equivalent width ($W$) between 
Fe~{\sc ii} 6147.7 and 6149.2 lines (which have almost the same 
strengths in the non-magnetic case; cf. Mathys 1990).\\ 
--- Successively, numerically solving the transfer equation of polarized 
radiation in the presence of a magnetic field according to Takeda (1991a),
Takeda (1991b; hereinafter referred to as T91b) explained the 
$H$-dependence of $W_{6147.7}/W_{6149.2}$ difference as due to the
desaturation effect caused by magnetic broadening
(while proposing the simultaneous use of similar line pair 
Fe~{\sc ii} 4416.8/4385.4), and derived
$H \sim$~2--3~kG (assuming a microturbulence of $v_{\rm t}\sim 0$~km~s$^{-1}$)
for $o$~Peg, which is favorably compared with ML90.\\
--- Further, Takeda (1993; hereinafter T93) devised a method for determining 
the magnetic field based on the $W$ values of many lines, which is regarded 
as a refined version of the classical Hensberge \& De~Loore's (1974) technique  
and can establish the ($H$, $v_{\rm t}$) solution by requiring the 
consistency of abundances derived from lines of various strengths.
Regarding $o$~Peg, T93 concluded $H\sim 2$~kG (and $v_{\rm t} \sim 1.5$~km~s$^{-1}$),
which is again consistent with ML90.

Since then, however, little progress seems to have been made as to the corroboration
of these findings. As already noted by ML90, the spectropolarimetric 
technique (most commonly used for investigating magnetic fields of CP stars) 
has been unsuccessful to accomplish a meaningful detection in $o$~Peg 
(though extremely weak magnetic field might as well be still detectable 
as remarked in footnote~1). 
Actually, besides the pioneering work of Babcock (1958),
Shorlin et al. (2002) could not detect any $\langle H_{z} \rangle$ 
(disk-averaged line-of-sight component of the field) of 
significance in $o$~Peg by their spectropolarimetric observation coupled with 
Least-Squares-Decomposition (LSD) technique.\footnote{
According to the compilation 
of Bychkov, Bychkova, \& Madej (2009), $o$~Peg's $\langle H_{z} \rangle$ 
resulting from Shorlin et al.'s (2002) observation was $32 (\pm 20)$~G, which
would have been regarded as under the level of significance.
}

This suggests that $\langle H_{z} \rangle$ happens to be too small 
(presumably because components of opposite signs are cancelled out
by disk integration) to produce a significant signal of circular polarization.

Accordingly, in order to study the magnetic nature of $o$~Peg, the best way would be 
to analyze the equivalent widths or line widths of many spectral lines 
(which contain information of ``absolute'' magnetic field strengths) as previously done.
However, it may be premature to regard the results of those old studies as sufficiently 
reliable because several problems are involved from a methodological point of view.  
\begin{itemize}
\item
The analysis of ML90 was not based on a rigorous modeling but a rather tentatively
postulated analytical relation between the line width, line strength, and 
Zeeman-broadening parameters. Especially, since the original Stenflo \& Lindegren's 
(1977) work (on which their study is based) was intended to estimate the solar 
magnetic field at the disk center, the effect of rotational broadening on 
the line width was not taken into account; thus, how the stellar projected rotational 
velocity ($v_{\rm e}\sin i$) affects the functional form of the relation is unclear.
\item
Although T91b simulated the emergent profiles and the strengths of spectral line 
pairs by correctly treating the transport of polarized radiation in a magnetic field,
only the specific intensity profiles at the disk center ($\mu = \cos\theta = 1$) 
were calculated for several single-valued field strengths ($H$)  and different angles 
between the magnetic field vector and the line of sight ($\psi$) , which must be 
unrealistic for comparing with the flux profiles of magnetic stars. Another 
problem is the choice of microturbulence: $v_{\rm t}\sim 0$~km~s$^{-1}$ adopted 
in T91b is not consistent with that obtained later by T93 ($\sim 1.5$~km~s$^{-1}$); 
if the latter were chosen, a considerably higher magnetic field ($\sim$~4--5~kG) 
would have resulted (cf. Sect.~4.1 in T91b). 
\item
The results of T93 were actually not robust but rather delicate, because $H$ 
could be firmly established only from Fe~{\sc ii} lines among the three species 
(finding a definite solution was difficult for Fe~{\sc i} and Ti~{\sc ii} lines).
Above all, the validity of the approximate method for evaluating the flux 
equivalent width under a magnetic field proposed by T93 (denoted as $W_{b}$,
which is a simple mean between the minimum and maximum intensification cases) 
should be quantitatively checked in the first place.
\item
Attention should also be paid to the adopted observational data of $o$~Peg. 
ML90's measurements of line widths for their application of Stenflo--Lindegren 
technique were done on spectra with a signal-to-noise ratio (S/N) of $\sim 100$ 
covering 3700--4650~\AA\ 
resulting from the co-added photographic spectrogram with a reciprocal 
dispersion of 2.4~\AA~mm$^{-1}$ obtained at Dominion Astrophysical Observatory 
(Adelman, Cowley, \& Hill 1988).   
Similarly, the equivalent width data of Fe~{\sc i}, Fe~{\sc ii}, and Ti~{\sc ii} 
lines for $o$~Peg adopted in the analysis of T93 were taken from Adelman's (1988) 
Table~5, for which he used the same co-added Dominion spectra. Since the quality 
of these photographic spectra is not sufficient from the viewpoint of present-day 
standard and using only the blue-region data is not advantageous for the purpose 
of magnetic-field detection, it is desirable to employ modern spectra of much 
higher quality covering wider (from blue through red) wavelength range.
\end{itemize}

As such, it is worth reinvestigating the magnetic field of $o$~Peg
by analyzing the widths and strengths for a number of spectral lines 
as done by ML90 and T93 but based on physically rigorous theoretical 
line-profile modeling along with new observational data.
Given this situation and now that CCD spectra of very high S/N  
for this star are now available, I decided to freshly revisit this problem, 
while making use of the profile simulation program based on disk integration 
coupled with the solution code of polarized radiation transfer in a magnetic field.
The purpose of this paper is to report the results of this new analysis.

%Sect. 2
\section{Observational data}

\subsection{Observed spectra of $o$~Peg}

The high-dispersion spectra of $o$~Peg adopted in this investigation are 
the same as used in Takeda et al.'s (2012) abundance study of alkali 
elements for A-type stars. The observations were done on UT 2008 October~4 
(8 frames of 1200~s exposure within a time span of $\sim$~6 hr), 
October~7 (1 frame of 1800~s exposure), October~8 (8 frames of 1200~s exposure 
within a time span of $\sim 3$~hr), and October~9 (6 frames of 1200~s exposure 
within a time span of $\sim 2$~hr) by using the HIDES (HIgh Dispersion Echelle 
Spectrograph) placed at the coud\'{e} focus of the 1.88~m reflector at Okayama 
Astrophysical Observatory. 
Equipped with three mosaicked 4K$\times$2K CCD detectors at the camera focus, 
echelle spectra covering 4100--7800~\AA\ (in the mode of red cross-disperser) 
with a resolving power of $R \sim 100000$ (corresponding to the slit width 
of 100~$\mu$m) were obtained. 

The reduction of the spectra (bias subtraction, flat-fielding, 
scattered-light subtraction, spectrum extraction, wavelength 
calibration, and continuum normalization) was performed by using 
the ``echelle'' package of the software IRAF\footnote{
IRAF is distributed by the National Optical Astronomy Observatories,
which is operated by the Association of Universities for Research
in Astronomy, Inc. under cooperative agreement with the National 
Science Foundation.} in a standard manner. 
In order to improve the signal-to-ratio, all available frames were co-added,
by which very high S/N (typically around $\sim 1000$ on the average) could be 
accomplished in the final spectra as shown in Fig.~1 (S/N $\sim \sqrt{count}$).   

%Fig. 1
\setcounter{figure}{0}
\begin{figure}[h]
\begin{minipage}{85mm}
\begin{center}
  \includegraphics[width=8.5cm]{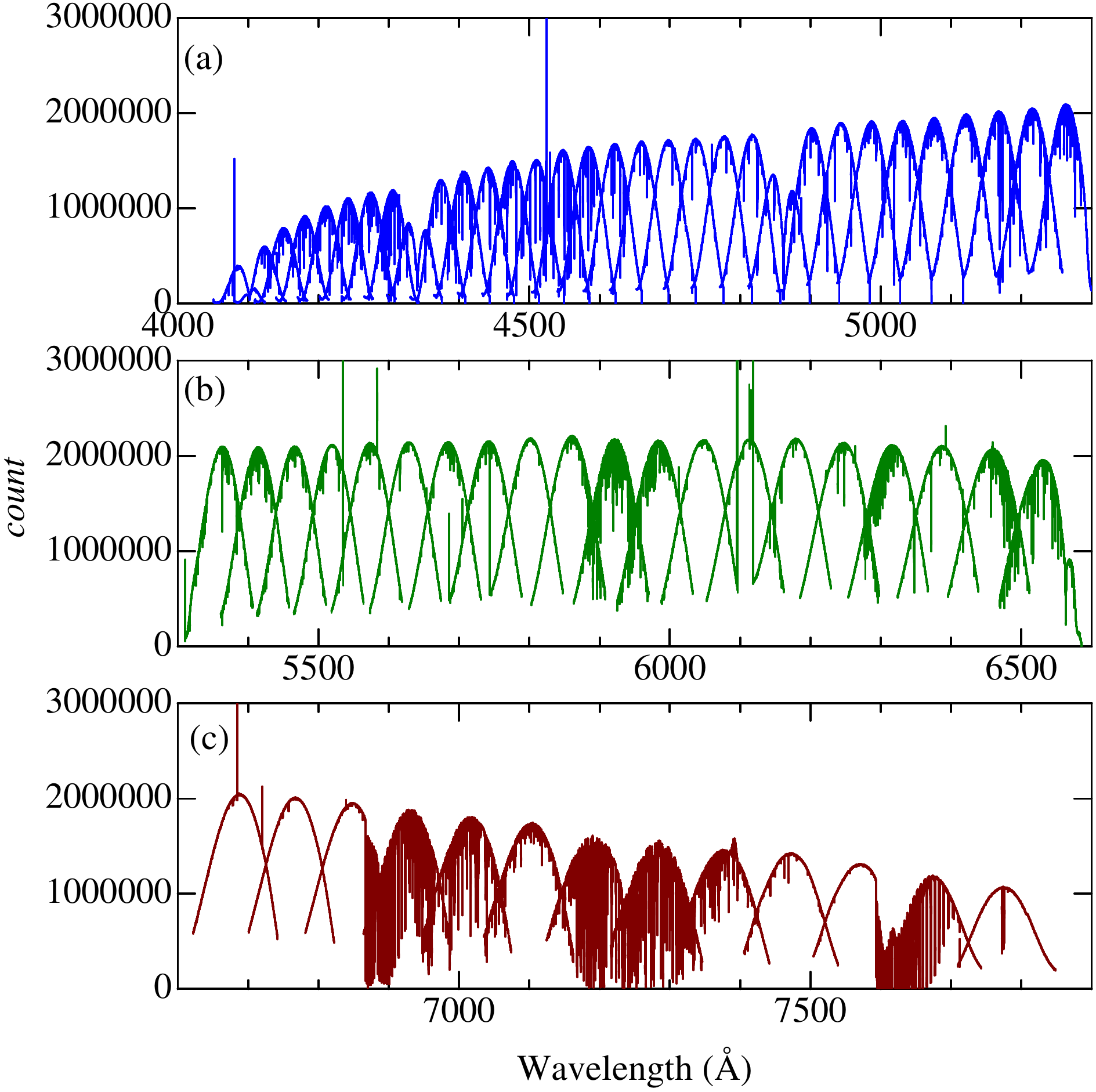}
\end{center}
\caption{
Panels (a), (b), and (c) show the distributions of accumulated photoelectron 
counts of CCD for the final spectra of $o$~Peg, each corresponding to three 
wavelength regions (4100--5300/5300--6600/6600--7900~\AA) comprising 32/20/13 
orders, respectively. Note that the signal-to-noise ratio can be estimated 
as S/N $\sim \sqrt{count}$ in the present photon-noise-limited case.
The spectra in each of the echelle orders show characteristic distributions of 
the blaze function.
}
\label{fig:1}
\end{minipage}
\end{figure}

\subsection{Selected lines and measurements}

In this study, we concentrate to using only spectral lines of Fe~{\sc i} and 
Fe~{\sc ii}, as they are available in larger number over a wide range of 
strengths than other species. The candidate lines to be analyzed were carefully 
sorted out by inspecting the observed spectral feature while comparing it 
with the calculated strengths of neighboring lines as well as the synthesized 
theoretical spectrum, as done by Takeda (2020). Because of the necessity of
calculating the Zeeman components, those lines lacking the information of
quantum numbers ($L$, $S$, $J$) for the lower and upper levels were excluded.
As a result,  198 Fe~{\sc i} and 182 Fe~{\sc ii} lines were selected.
The equivalent widths ($W_{\lambda}$) of these lines were measured by
fitting the line-depth profile ($R_{\lambda} \equiv 1 - f_{\lambda}/f_{\rm cont}$) 
with the Gaussian function ($\propto \exp [-(\lambda - \lambda_{0})^{2}/a^{2}]$), 
while the line widths ($h_{\lambda}$; defined as the FWHM of $R_{\lambda}$) 
were directly evaluated from the profiles.

The $W_{\lambda}$ values of these Fe lines range from $\sim 1$~m\AA\ to
$\sim 200$~m\AA\ (Fig.~2a) and their effective Land\'{e} factors are between
$0 \lesssim g_{\rm eff}^{\rm L} \lesssim 3$ (Fig.~2b).
As seen from the empirical curves of growth depicted in Fig.~2c,
the linear part and the shoulder/flat part are roughly separated 
around $10^{6} W_{\lambda}/\lambda \sim 10$ (typically several tens 
m\AA\ in $W_{\lambda}$).  
A comparison of the equivalent widths with those published by Adelman (1988) 
is displayed in Fig.~2d, where a reasonable consistency is observed.  
Fig.~2e shows that the directly measured FWHM values ($h_{\lambda}$) are  
mostly in agreement with the corresponding Gaussian-fit ones (derived from 
the $e$-folding half-width as $h_{\lambda}^{\rm G.F.} \equiv 2\sqrt{\ln 2} \; a$).
The dependence of $h_{v}$ (in the velocity unit derived as 
$h_{\lambda} c/\lambda$: $c$ is the speed of light) 
upon the equivalent width is illustrated in Fig.~2f, 
which shows that $h_{v}$ begins to exhibit a systematic 
$W_{\lambda}$-dependence at $10^{6} W_{\lambda}/\lambda \gtrsim 10$. 

The measured $W$ and $h$ values of these 380 Fe lines along with their 
atomic data (wavelength, excitation potential, oscillator strengths,
damping constants, term information, effective Land\'{e} factor, etc.) 
mostly taken from the VALD database (Ryabchikova et al. 2015)  
are summarized in ``felines.dat'' of the online material,
where the original profile data of all lines are also given  
in ``obsprofiles.dat''.

%Fig. 2
\setcounter{figure}{1}
\begin{figure*}[h]
\begin{minipage}{180mm}
\begin{center}
  \includegraphics[width=12cm]{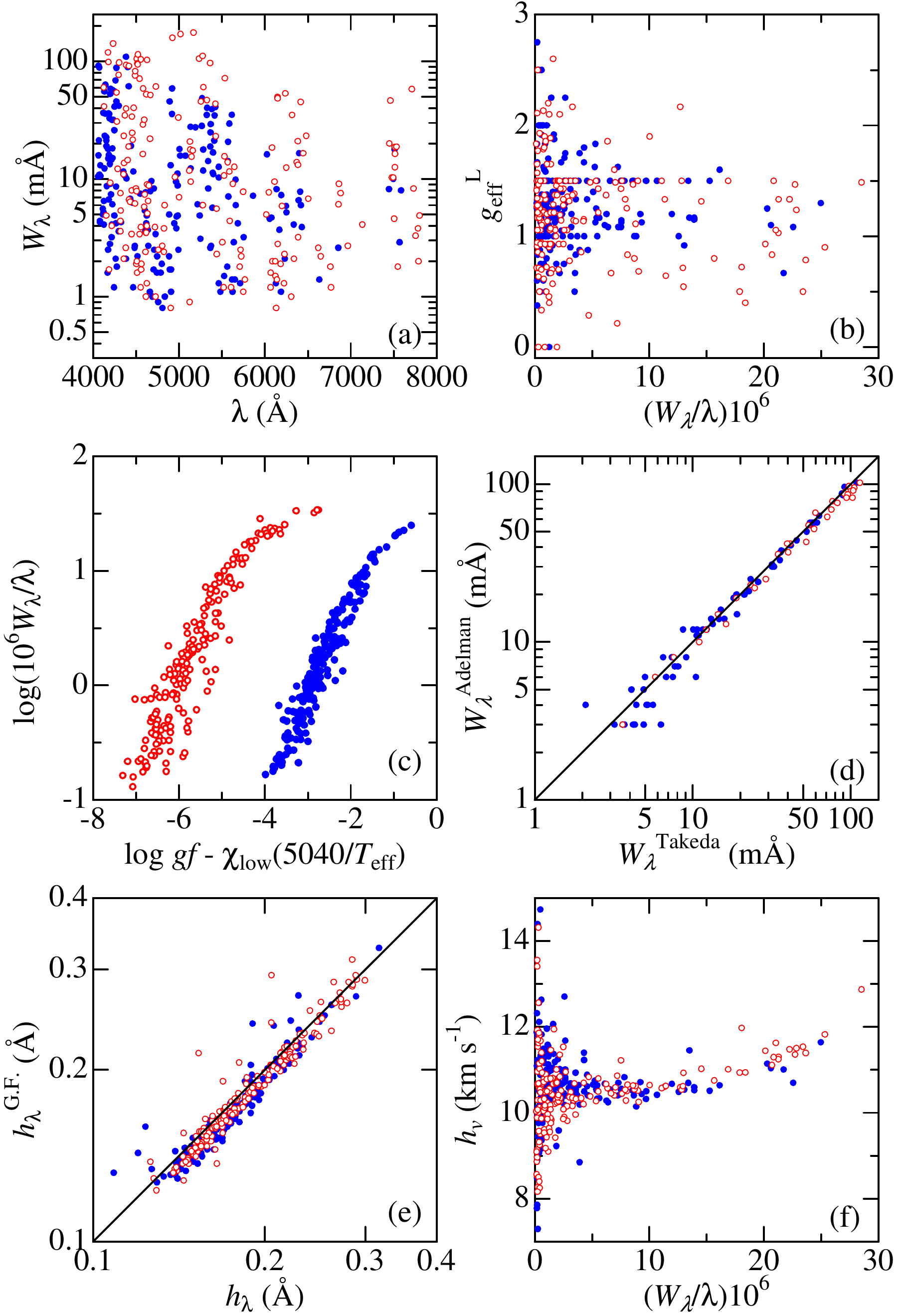}
\end{center}
\caption{
(a) Equivalent width vs. wavelength.
(b) Effective Land\'{e} $g$ factor vs. reduced equivalent width.
(c) Empirical curves of growth for the Fe~{\sc i} and Fe~{\sc ii} lines,
where $\log gf - \chi_{\rm low}(5040/T_{\rm eff})$ is taken in the abscissa
($g$: statistical weight of the lower level, $f$: oscillator strength,
$\chi_{\rm low}$: lower excitation potential in eV, $T_{\rm eff}$: effective
temperature in K).
(d) Comparison of the equivalent widths measured in this study 
with those published by Adelman (1988) for 97 lines (65 Fe~{\sc i} lines
and 32 Fe~{\sc ii} lines) in common. 
(e) Correlation between the directly measured full-widths at half-maximum 
($h_{\lambda}$) with the corresponding Gaussian-fit values 
($h_{\lambda}^{\rm G.F.}$).
(f) Full-widths at half-maximum in the velocity unit 
($h_{v} \equiv c h_{\lambda}/\lambda$; $c$ is the speed of light)
plotted against the reduced equivalent widths.
In each panel, the data for Fe~{\sc i} and Fe~{\sc ii} lines are discriminated
in blue filled symbols and red open symbols, respectively. 
}
\label{fig:2}
\end{minipage}
\end{figure*}

% Table 1
\setcounter{table}{0}
\begin{table*}[h]
\begin{minipage}{180mm}
\caption{Atmospheric parameters of $o$ Peg published so far.}
%\small
\scriptsize
\begin{center}
\begin{tabular}{ccccccl} 
\hline\hline
Authors & $T_{\rm eff}$ & $\log g$ & $v_{\rm t}$ & [Fe/H] & $v_{\rm e}\sin i$ \\
\hline
Wolff  (1967) & 9330 & 3.2  & 2.0  & 0.30  &  \\
Conti \& Strom (1968) & 9500 & 4.0  & 3.0  & 0.2$^{a}$  & \\
Adelman (1973) & 10100 & 4.0  & 3.0  & 0.1  &  \\
Allen (1977) & 9600 & 3.8  & 3.2  & $-0.09$  &  \\
Mitton (1977) & 9500 & 3.5  & 1.6  & 0.08$^{b}$  &  \\
Adelman et al. (1984) & 9625 & 3.45  & 1.6  & 0.20  & \\
Adelman \& Fuhr (1985) &  &  & 1.9  & 0.26  &  \\
Adelman (1988) & 9600 & 3.60  & 1.3  & 0.10  & 6  \\
Castelli \& Hack (1988) & 9590 & 3.55  & 1.9  & 0.14  & 6  \\
Kocer et al. (1988) & 9500 & 3.50  & 1.5  & 0.04  &  \\
Sadakane (1988) & 9500 & 3.50  & 2.0  & 0.02  &  \\
Van't Veer et al. (1988) & 9350 & 3.50  & 1.5  & 0.02  &  \\
Burkhart \& Coupry (1991) & 9650 & 3.6  &   & 0.1  &  \\
Hill \& Landstreet (1993) & 9680 & 3.71  & 1.5$^{c}$ & 0.03  & \\
Abt \& Morrell (1995) &  &  &  &  & 10  \\
Hill (1995) &  &  & 1.7  & 0.19  & 6.3  \\
Sokolov (1995) & 10050 &  &  &  &  \\
Blackwell \& Lynas-Gray (1998) & 9443 &  &  &  &  \\
Di Benedetto (1998) & 9720 &  &  &  & \\
Hui-Bon-Hoa (2000) & 9650 & 3.6  & 1.5  & 0.42  & $\le 10$ \\
Adelman et al. (2002) & 9591 & 3.64  &  &  & \\
Adelman et al. (2002) & 9525 & 3.70  &  &  & \\
Royer et al. (2002) &  &  &  &  & 14  \\
Royer et al. (2007) &  &  &  &  & 14  \\
Landstreet et al. (2009) & 9500 & 3.62  & 2.0  & 0.14$^{d}$  & 7  \\
Zorec et al. (2009) & 9930 &  &  &  &  \\
Prugniel et al. (2011) & 9373 & 3.73  &  & $-0.14$  &  \\
Takeda et al. (2012) & 9453 & 3.64  & 3.1  & 0.13$^{e}$  & 6.0  \\
Zorec \& Royer (2012) & 9506 & 3.73$^{f}$  &  &  & 14 \\
Gray (2014) & 9600 & 3.7  &  & 0.0$^{g}$  & 6.00$^{h}$ \\
Takeda et al. (2018) & 9453 & 3.64  & 2.7  & 0.18$^{i}$  & 6.6  \\
\hline
\end{tabular}
\end{center}
Summarized here are the effective temperature (in K), logarithmic surface gravity
in c.g.s unit (in dex), microturbulence (in km~s$^{-1}$), Fe abundance relative to the Sun, 
and projected rotational velocity (in km~s$^{-1}$) of $o$~Peg taken from previous publications. 
Since these parameters were determined in variously different methods, the original references 
should be consulted for the details. Regarding [Fe/H], if only those derived from Fe~{\sc i}
and Fe~{\sc ii} lines are available, a simply averaged value of these two is listed here. 
Besides, in case that the reference solar Fe abundance is not explicitly given, an appropriate value
widely used at the time of publication was tentatively adopted.\\ 
$^{a}$Relative to the mean of 4 standard stars.\\
$^{b}$Relative to Procyon.\\
$^{c}$Assumed.\\
$^{d}$Solar Fe abundance of $\log({\rm Fe/H})_{\odot} = -4.49$ was assumed.\\
$^{e}$Solar Fe abundance of $A_{\odot}$(Fe) = 7.50 was assumed.\\
$^{f}$Derived from $L$ (bolometric luminosity), $T_{\rm eff}$, and $M$ (mass).\\
$^{g}$Assumed.\\
$^{h}$Radial-tangential macroturbulence of $\zeta_{\rm RT} = 5.7$~km~s$^{-1}$ was adopted.\\
$^{i}$Relative to Procyon.
\end{minipage}
\end{table*}

%Sect. 3
\section{Conventional analysis of equivalent widths for microturbulence}

Before dealing with the main issue of magnetic field estimation to be 
described in Sect. 4 and 5, we first conduct a preparatory analysis of 
determining the microturbulence based on the equivalent widths by applying 
the conventional procedure (while assuming as if no magnetic field exists).

\subsection{Atmospheric model and parameters}

Regarding the standard model atmosphere of $o$~Peg, we adopted Kurucz's (1993)
ATLAS9 solar abundance model with $T_{\rm eff} = 9500$~K, $\log g = 3.60$ 
(cgs unit).
These atmospheric parameters were chosen by inspecting the various literature 
data summarized in Table~1. As recognized from this table, these values 
(though rather rounded) are consistent with those derived 
in many of the past studies (especially the recent ones published after 2000).
Besides, this choice is in accord with $M = 2.8 M_{\odot}$ (mass) and 
$R = 4.4 R_{\odot}$ (radius) evaluated from the position on the
$\log L$ (luminosity) vs. $T_{\rm eff}$ diagram in comparison with 
theoretical evolutionary tracks (see, e.g.,  Fig.~1 in Takeda et al. 2012). 
Abundance determination from an equivalent width for a given microturbulence  
was done by using Kurucz's (1993) WIDTH9 program while assuming LTE. 
The adopted line parameters are given in ``felines.dat'' (see Sect.~2.2).  

Two approaches are tried in this $v_{\rm t}$ determination test; (1) usual 
method of finding the minimum abundance dispersion, and (2) alternative method 
requiring the overall consistency between the observed and theoretical $W$.  

\subsection{Method 1: minimum abundance dispersion}

The effect of $v_{\rm t}$ on abundance determination appreciably depends 
on line strengths: abundances determined from weak lines in the linear part 
of the curve of growth are essentially free from $v_{\rm t}$, while those 
from stronger lines on the shoulder-to-flat part are considerably 
$v_{\rm t}$-dependent. Therefore, $v_{\rm t}$ is usually established by 
requiring that abundances derived from lines of various equivalent widths 
($W$) be consistent with each other.

Among the several practical procedures for accomplishing this requirement, 
Blackwell et al.'s (1976) method is used here.\\
--(1) For each line $n$, a set of abundances  ($A_{n}^{k}$;  
$k = 1, 2,\ldots, K$) are derived from $W_{n}$ while incrementally 
changing the microturbulence ($v_{\rm t}^{k}$; $k = 1, 2,\ldots, K$).\\
--(2) Then, the mean abundance ($\langle A \rangle^{k}$)
averaged over $N$ lines and the standard deviation $\sigma_{A}^{k}$ 
are calculated for each of the $K$ microturbulences ($v_{\rm t}^{k}$).\\
--(3) By inspecting the resulting standard deviation ($\sigma_{A}^{k}$; $k = 1, 2,\ldots, K$),
the location of minimum $\sigma_{A}$  corresponds to desired solution of $v_{\rm t}$.

This procedure was applied to our data set of Fe lines. 
The resulting $A$ vs. $v_{\rm t}$ relations for each of the lines 
along with the corresponding $\sigma_{A}$ vs. $v_{\rm t}$ curve
are shown in Figs.~3a and 3b for Fe~{\sc i} ($N_{1} = 198$) and
Fe~{\sc ii} lines ($N_{2} = 182$), respectively.
As seen from these figures, we obtain (1.71~km~s$^{-1}$, 7.75)\footnote{
$A$ is the logarithmic number abundance of Fe relative to H,
which is expressed in the usual normalization of 
$\log [N({\rm Fe})/N({\rm H})] + 12$.}
 and (1.76~km~s$^{-1}$, 7.82) as the results of ($v_{\rm t}$, $\langle A \rangle$)
for Fe~{\sc i} and Fe~{\sc ii}; and the corresponding $A_{n}$ for each line 
is plotted against $W_{n}$ in Figs.~3c and 3d, respectively .

%Fig. 3
\setcounter{figure}{2}
\begin{figure}[h]
\begin{minipage}{85mm}
\begin{center}
  \includegraphics[width=8.5cm]{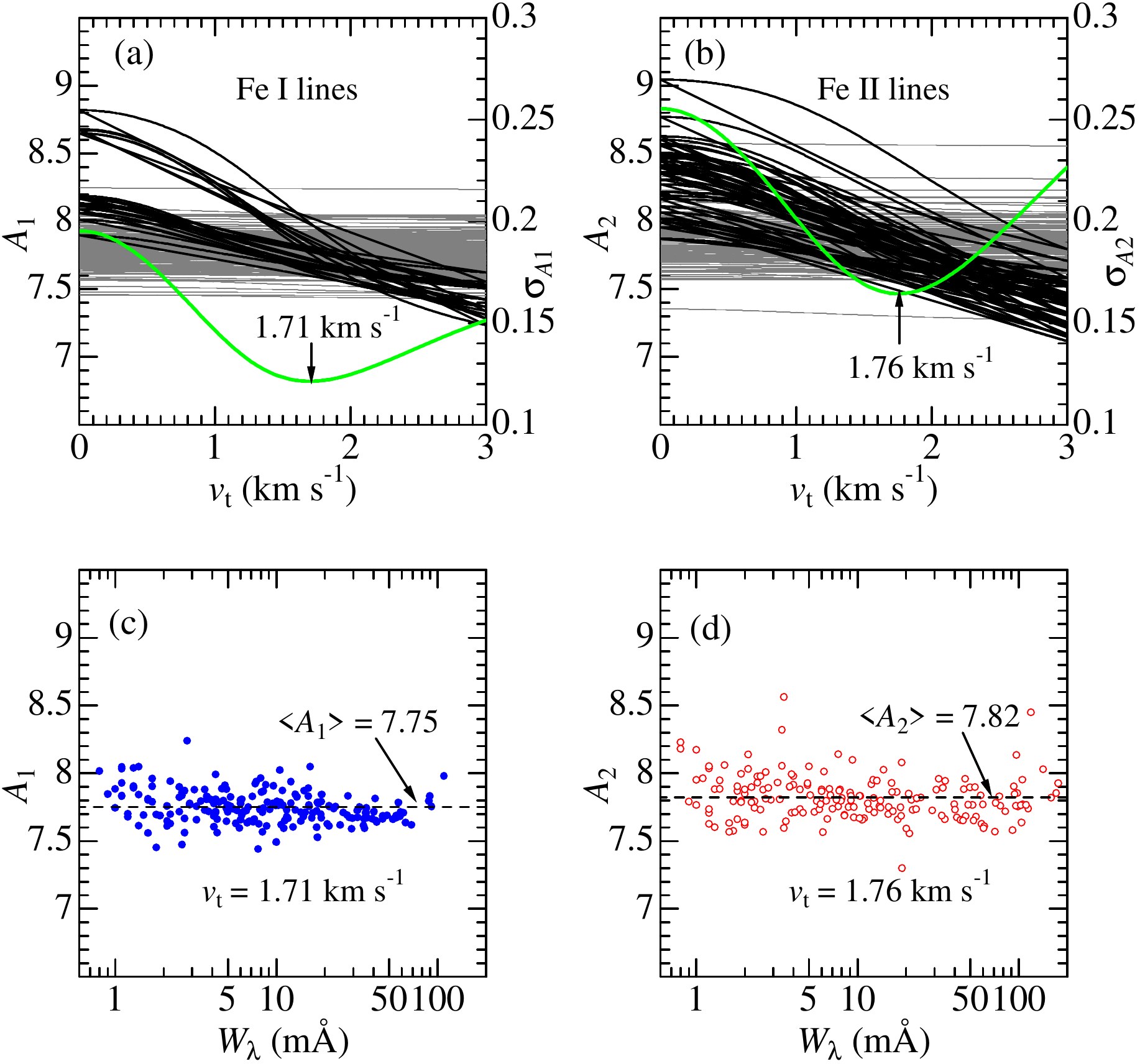}
\end{center}
\caption{
Upper panels (a), (b): solid lines show how the Fe abundance ($A$) of 
each line varies by changing $v_{\rm t}$, where weaker lines 
($10^{6} W_{\lambda}/\lambda  < 10$) and stronger lines 
($10^{6} W_{\lambda}/\lambda > 10$), are distinguished by
gray and black lines, respectively. 
In addition, $\sigma_{A}$ (standard deviation of $A$) is plotted 
against $v_{\rm t}$ by the thick green solid line (its scale is marked 
in the right axis), and the $v_{\rm t}$ solution corresponding to 
the minimum $\sigma_{A}$) is also indicated. 
Lower panels (c), (d): Fe abundances corresponding to
the $v_{\rm t}$ solution are plotted against
the observed equivalent widths. The mean abundance ($\langle A \rangle$)
is also indicated by the horizontal dashed line.
The left-hand and right-hand panels are for Fe~{\sc i} 
and Fe~{\sc ii} lines, respectively.
}
\label{fig:3}
\end{minipage}
\end{figure}

\subsection{Method 2: minimum equivalent width dispersion}

Next, another method for $v_{\rm t}$ determination is tried. Since 
$A = 7.8$ may be regarded as the fiducial Fe abundance of $o$~Peg
from the results of 7.75 (Fe~{\sc i}) and 7.82 (Fe~{\sc ii}) derived in 
the previous subsection, we can calculate the theoretical equivalent widths 
($W_{{\rm cal},n}^{k,7.8}$; $n=1, 2, \ldots, N$) for each of the lines 
with this {\it fixed} Fe abundance but incrementally changing $v_{\rm t}^{k}$, which 
are to be compared with the observed ones ($W_{{\rm obs},n}$; $n=1, 2, \ldots, N$).
Then, by examining the standard deviations $\sigma_{W}^{k}$ evaluated for 
various $v_{\rm t}^{k}$ values,
\begin{equation}
\sigma_{W}^{k} \equiv 
\sqrt{\sum_{n=1}^{N}(W_{{\rm cal},n}^{k,7.8} - W_{{\rm obs},n})^{2}/N},
\end{equation}
we can find the solution of $v_{\rm t}$ as that accomplishing 
the minimum $\sigma_{W}^{k}$. 
Although the final $W_{\rm cal}$ calculated with such determined $v_{\rm t}$  
is not exactly equal to the observed $W_{\rm obs}$ (because of the 
fixed Fe abundance), the primary aim of line-independent consistency 
(i.e., without any global $W$-dependent trend) can be accomplished.  

The differences $W_{{\rm cal},n}^{k,7.8} - W_{{\rm obs},n}$ for each lines
are plotted against $v_{\rm t}$ in Fig.~4a (Fe~{\sc i}) and 4b (Fe~{\sc ii}),
where the corresponding $\sigma_{W}$ versus $v_{\rm t}$ curves are also depicted.
From these figures, we obtain 1.38~km~s$^{-1}$ and 1.77~km~s$^{-1}$ for 
Fe~{\sc i} and Fe~{\sc ii}, respectively. 
The differences $W_{{\rm cal},n}^{k,7.8} - W_{{\rm obs},n}$ for each lines 
corresponding to these $v_{\rm t}$ solutions are plotted against 
$W_{{\rm obs},n}$ in Fig.~4c (Fe~{\sc i}) and 4d (Fe~{\sc ii}),

A comparison of $v_{\rm t}$(Method 1) derived in Sect.~3.2 with this 
$v_{\rm t}$(Method 2) suggests that, while we can confirm a good agreement 
for the case of Fe~{\sc ii} lines (1.76/1.77~km~s$^{-1}$), a discrepancy 
is seen for $v_{\rm t}$ based on Fe~{\sc i} lines (1.71/1.38~km~s$^{-1}$).
As a matter of fact, the results from Fe~{\sc i} lines appear to be
somewhat problematic. As can be seen in Fig.~3c, the distribution of 
$W_{{\rm cal},n}^{k,7.8} - W_{{\rm obs},n}$ differences for Fe~{\sc i} lines 
of medium-to-large strengths ($W_{\rm obs} \gtrsim 10$~m\AA)
shows some asymmetric feature (i.e., positive for lines of 20--50~m\AA\
while negative for those of $\gtrsim 50$~m\AA). This trend has made 
the $\sigma_{W}^{k}$ curve shallower with a less clear minimum (Fig.~4a), 
which eventually leads to larger uncertainties in $v_{\rm t}$ determination.
Generally speaking, since only a very tiny fraction of Fe atoms remain neutral 
(those in Fe~{\sc ii} and Fe~{\sc iii} stages are dominant) in the atmosphere 
of A-type stars, the formation of Fe~{\sc i} lines is considerably $T$-dependent
and vulnerable to model atmosphere structure, while Fe~{\sc ii} lines are 
more robust in this respect (see Appendix~A2 of Takeda 2020). Accordingly, 
Fe~{\sc ii} lines may yield more reliable results than Fe~{\sc i} lines. 

%Fig. 4
\setcounter{figure}{3}
\begin{figure}[h]
\begin{minipage}{85mm}
\begin{center}
  \includegraphics[width=8.5cm]{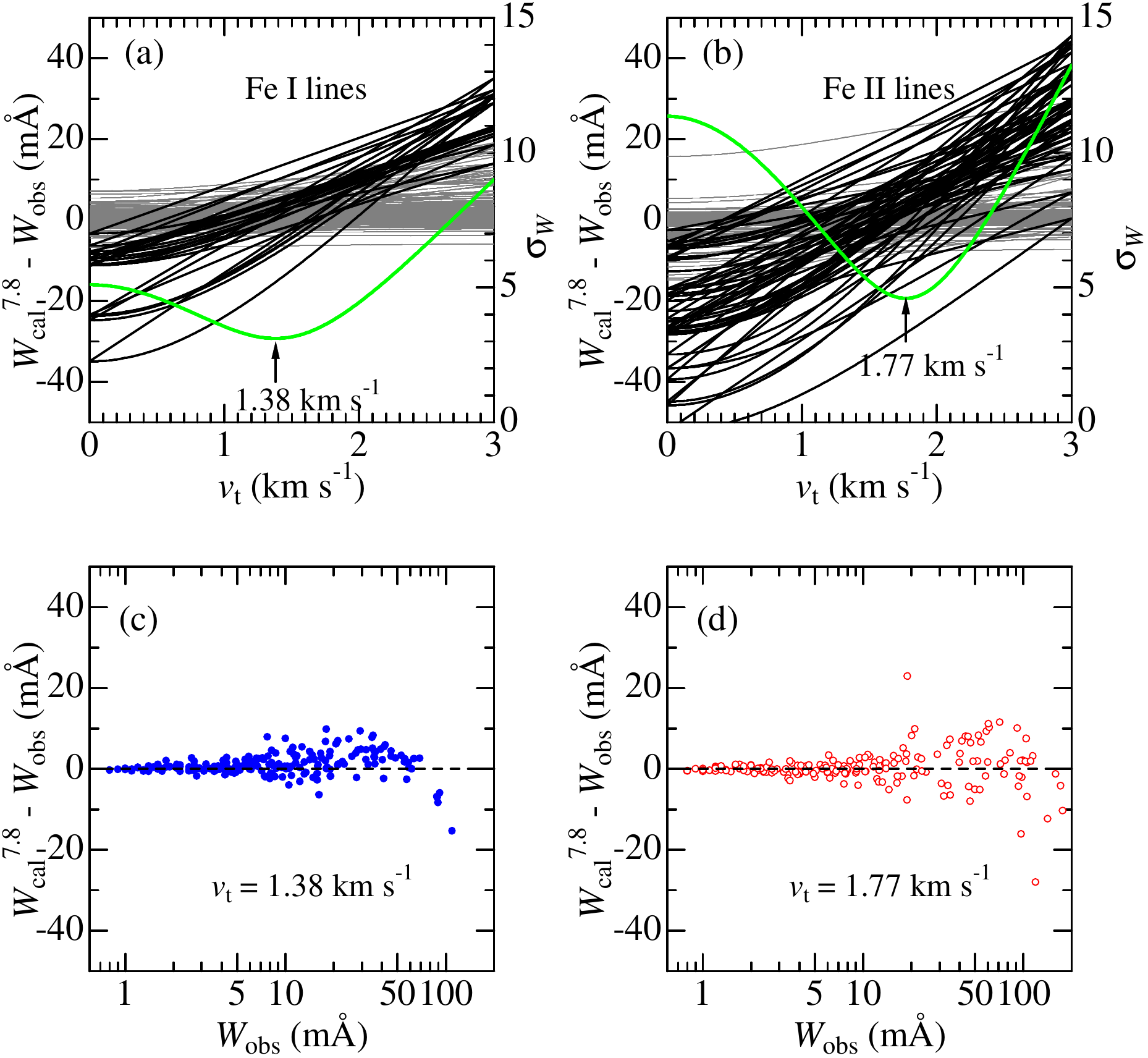}
\end{center}
\caption{
Upper panels (a), (b): solid lines show how the difference between 
the theoretical equivalent width calculated for $A = 7.8$ ($W_{\rm cal}^{7.8}$)
and the observed equivalent width ($W_{\rm obs}$) varies 
by changing $v_{\rm t}$, where weaker and stronger lines are distinguished 
by gray and black lines, respectively (as in Fig.~3).
The standard deviation ($\sigma_{W}$) of the differences defined by Eq.~(1)  
is depicted against $v_{\rm t}$ in thick green line (scale is in the right axis).
Lower panels (c), (d): Equivalent width differences
($W_{\rm cal}^{7.8} - W_{\rm obs}$) corresponding to the $v_{\rm t}$ 
value of minimum $\sigma_{W}$ are plotted against $W_{\rm obs}$.
The left-hand and right-hand panels are for Fe~{\sc i} 
and Fe~{\sc ii} lines, respectively.
}
\label{fig:4}
\end{minipage}
\end{figure}

%Sect. 4
\section{Line profile simulation of a magnetic star}

\subsection{Magnetic field model}

As already mentioned in Sect.~1, the main aim of this investigation is 
to check the previously reported results (possible existence of a magnetic 
field on the order of $\sim 2$~kG in $o$~Peg) based on physically legitimate 
simulations of Zeeman-broadened line profiles.
What matters here is the choice of rotating magnetic star models (field 
configuration,inclination of magnetic/rotational axes viewed by an 
observer, etc.) among diversified possibilities. 
In the context of lacking information, we tentatively assume a model 
which is as simple as possible but does not yield results contradicting 
the known observational facts.
In any case, given that we are primarily interested in the value of 
$\langle H \rangle$ (mean field strength averaged over the disk; cf. Eq.~(3)),
we do not need to be too much particular about this issue, because the functionality 
of $h(\langle H\rangle ,v_{\rm e}\sin i)$ or $W(\langle H\rangle ,v_{\rm t})$ 
would not be very sensitive to any choice of models in the first approximation, 

Following this policy, a simple dipole model is adopted
in this study, which is represented in the spherical coordinate system
as follows: 
\begin{eqnarray}
H_{r} & = & H_{\rm pol} (R/r)^{3} \cos\theta \nonumber \\
H_{\theta} & = & (H_{\rm pol}/2) (R/r)^{3} \sin\theta \nonumber \\
H_{\phi} & = & 0,
\end{eqnarray}
where $H_{\rm pol}$ is the magnetic field strength 
at the magnetic pole on the stellar surface 
($r=R$, $\theta=0^{\circ}$),
which yields $(H_{r}, H_{\theta}, H_{\phi})$ = 
$(H_{\rm pol} \cos\theta, H_{\rm pol}/2\, \sin\theta, 0)$
at the stellar surface ($r = R$).
Accordingly, the absolute strength of the surface field is largest at the 
magnetic pole ($\theta = 0^{\circ}$, $|{\bf H}| = H_{\rm pol}$) and smallest 
at the equator ($\theta = 90^{\circ}$, $|{\bf H}| = H_{\rm pol}/2$)

As to the axis orientation and view angle of this model star, we assume that 
the magnetic and rotational axes are in line with each other and perpendicular 
to the observer's line of sight; i.e., $i = \alpha = 90^{\circ}$ (as usual, 
$i$ and $\alpha$ are the angles of rotational and magnetic axes in reference 
to the line of sight), as shown in the upper illustration of Fig.~5. 
This simple assumption is reasonable in context of the observational characteristics
of $o$~Peg, because (i) the magnetic field configuration viewed by the observer does not 
depend upon the rotational phase (i.e., no appreciable variability) and (ii) the 
line-of-sight component of the field is cancelled out by averaging over the disk to 
result in $\langle H_{z} \rangle =0$ (meaningful circular polarization signal is undetected). 
The observed aspect of surface magnetic field in this model is schematically 
depicted in the lower-left panel (meridional cross section) and lower-middle 
panel (observer's view) of Fig.~5. 
Besides, the magnetic field strengths $|{\bf H}|$ (in unit of $H_{\rm pol}$) 
at various points on the visible disk are plotted against $\cos \psi$ ($\psi$ 
is the angle between the field vector and the line of sight) in the lower-right 
panel of Fig.~5, from which we can see that $|{\bf H}|$ is between $H_{\rm pol}/2$ 
and $H_{\rm pol}$.

%Fig. 5
\setcounter{figure}{4}
\begin{figure*}[h]
\begin{minipage}{180mm}
\begin{center}
  \includegraphics[width=12cm]{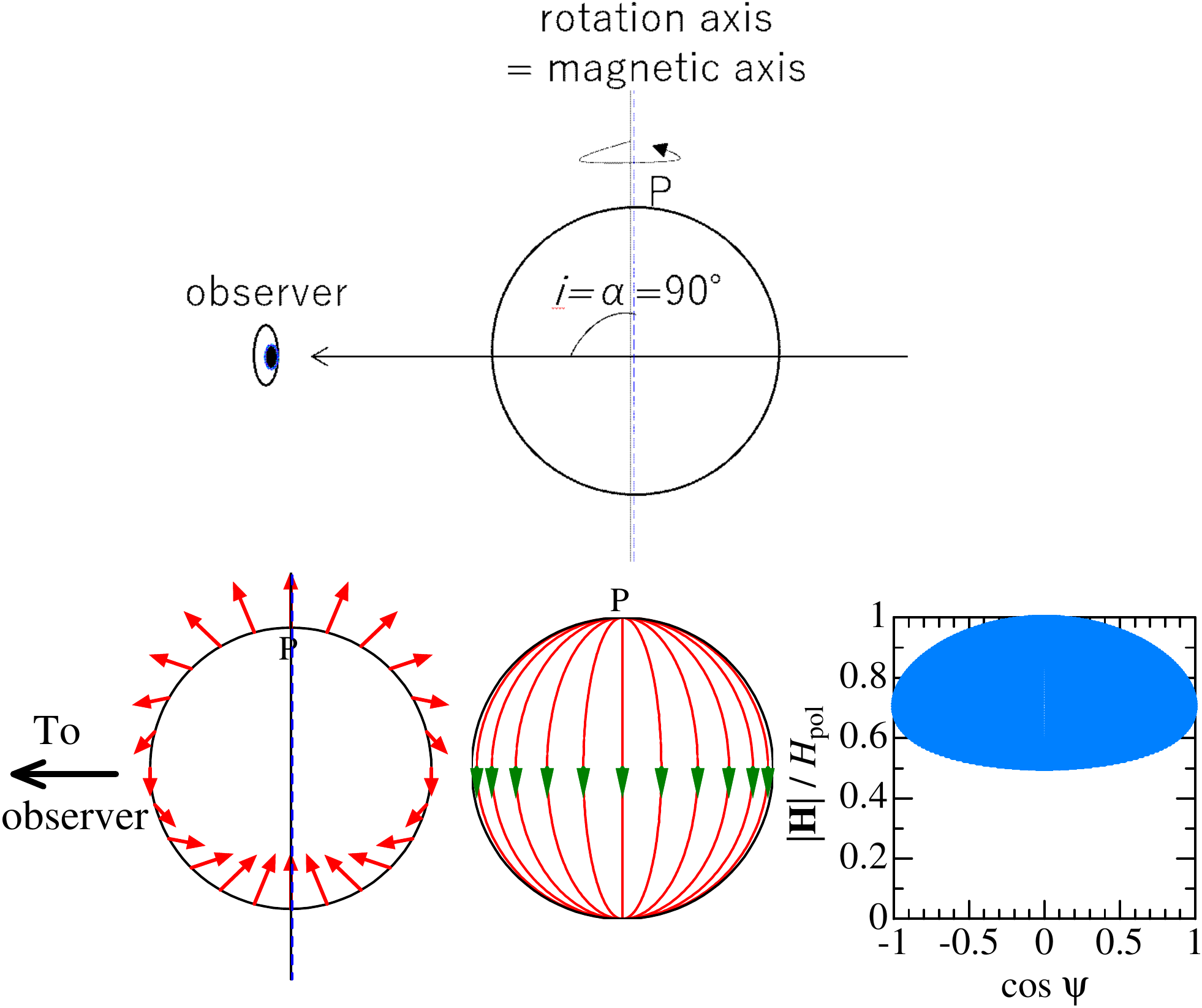}
\end{center}
\caption{
The upper figure schematically describes the adopted rotating star model with a 
dipole magnetic field  (parameterized by $H_{\rm pol}$; field strength at the pole P), 
where the observer's line of sight is perpendicular to both of the rotational 
axis ($i = 90^{\circ}$) and the magnetic axis ($\alpha = 90^{\circ}$).
The lower three figures represent the observed characteristics of the magnetic field
in this model:  Left --- surface field vectors (indicated by arrows) in the meridional 
plane. Center --- schematic illustration of surface magnetic field lines viewed 
by an observer. Right --- Correlation between $|{\bf H}|/H_{\rm pol}$ 
(absolute field strength in unit of $H_{\rm pol}$ and $\cos\psi$ ($\psi$ is the angle 
between the magnetic field vector and the line of sight) at each point of the visible disk.
}
\label{fig:5}
\end{minipage}
\end{figure*}

\subsection{Calculation of local $I$ profiles}

As to the calculation of specific intensity profile ($I_{\lambda}$) of a line 
under the presence of a magnetic field, Unno's (1956) radiative transfer 
equation in terms of the Stokes parameters ($I, Q, V$) was solved with 
the help of Takeda's (1991a) numerical procedure (cf. Sect.~2 therein) 
as done by T91b.
The line opacity profiles of Zeeman-split $\pi$, $\sigma_{+}$, 
and $\sigma_{-}$ components (derived from $S$, $L$, and $J$ of
upper and lower levels by assuming LS coupling) for a given magnetic field 
were evaluated by making use of the line opacity data of non-magnetic case 
calculated by the WIDTH9 program (with the same model atmosphere as adopted in Sect.~3).
The necessary parameters for computing the $I_{\lambda}$ profile 
emergent from a disk point are $|{\bf H}|$, $\psi$, $\mu$ (direction cosine 
of the angle between the surface normal and the line of sight), along with 
$A$ (Fe abundance) and $v_{\rm t}$ (microturbulence).
Accordingly, a grid of emergent $I_{\lambda}^{\rm grid}$ profiles (up to 
1\AA\ from the line center with a step of 0.005\AA) were computed in advance 
for each of the 380 lines for combinations of
31 $|{\bf H}|$ values (0, 200, 400, $\ldots$ 5800, 6000~G),
10 $\psi$ values (0, 10, 20, $\ldots$, 80, 90$^{\circ}$),
10 $\mu$ values (0.1, 0.2, 0.3, $\cdots$, 0.9, 1.0), and
7 $v_{\rm t}$ values (0.0, 0.5, 1.0, $\ldots$, 2.5, 3.0~km~$^{-1}$),
while the Fe abundance was fixed at $A = 7.80$ (cf. Sect. 3).

\subsection{Line flux profile by disk integration}

The flux profile $F_{\lambda}$ of a spectral line can then be simulated
by integrating the $I_{\lambda}$ at each point of the visible disk 
(evaluated by interpolating the grid of $I_{\lambda}^{\rm grid}$
corresponding to the local physical condition), while adequately taking 
into account the Doppler shift due to the line-of-sight velocity.
For this purpose, we modified the program CALSPEC (Takeda, Kawanomoto, \& Ohishi 2008)
which simulates the spectral line profile of a rotating star by dividing 
its surface into 180$\times$360 segments. Since only the case of slow rotation 
is considered, the effects of gravity darkening and gravitational distortion
were neglected; therefore, the star is spherical and homogeneously covered 
with the solar abundance atmosphere of $T_{\rm eff} = 9500$~K and $\log g = 3.60$.
The parameters of $H_{\rm pol}$ and $v_{\rm e}\sin i (= v_{\rm e})$ 
have to be assigned (along with $v_{\rm t}$ and $A$) in this modeling of 
line flux profile.

The calculations of $F_{\lambda}$ for each line were done for  
13 $H_{\rm pol}$ values (0, 500, 1000, $\ldots$ 5500, 6000~G),  
7 $v_{\rm e}\sin i$ values (0.0, 2.5, 5.0, $\ldots$, 12.5, 15.0~km~s$^{-1}$), 
and 7 $v_{\rm t}$ values (0.0, 0.5, 1.0, $\ldots$, 2.5, 3.0~km~$^{-1}$),
again at the fixed $A = 7.80$.
Further, the equivalent widths ($W_{\rm cal}$) and FWHMs 
($h_{\rm cal}$) were also evaluated from these line profiles. 
As an example of simulation, the $F_{\lambda}$ results derived for 
representative three lines (Fe~{\sc i} 4383.544, Fe~{\sc ii} 6147.734, and
Fe~{\sc ii} 6149.246) are displayed in Fig.~6, where the corresponding observed
profiles are also shown for comparison.

For the sake of future discussion, the mean absolute field strength 
averaged over the visible stellar disk ($\langle H \rangle$) is defined as follows:
\begin{equation}
\langle H \rangle \equiv \left.
\int_{\rm disk}\!\!\!\int |{\bf H}|(x,y) I_{\rm cont}(x,y) {\rm d}x{\rm d}y 
\middle/
\int_{\rm disk}\!\!\!\int I_{\rm cont}(x,y) {\rm d}x{\rm d}y\right. ,
\end{equation}
where $|{\bf H}|(x,y)$ and $I_{\rm cont}(x,y)$ are the absolute field strength and
the continuum specific intensity (to the observer) at the disk point $(x, y)$, 
respectively.
Naturally, $\langle H \rangle$ is in proportion to $H_{\rm pol}$ with the proportionality 
constant of $\langle H \rangle / H_{\rm pol}$ = 0.642 in the postulated magnetic 
field configuration ($\alpha = 90^{\circ}$). 
Likewise, the disk-averaged line-of-sight component (in the $z$-direction) 
of the magnetic field ($\langle H_{z} \rangle$) is definable in the similar manner 
and $\langle H_{z} \rangle = 0$ holds in the present case. 

%Fig. 6
\setcounter{figure}{5}
\begin{figure*}[h]
\begin{minipage}{180mm}
\begin{center}
  \includegraphics[width=12cm]{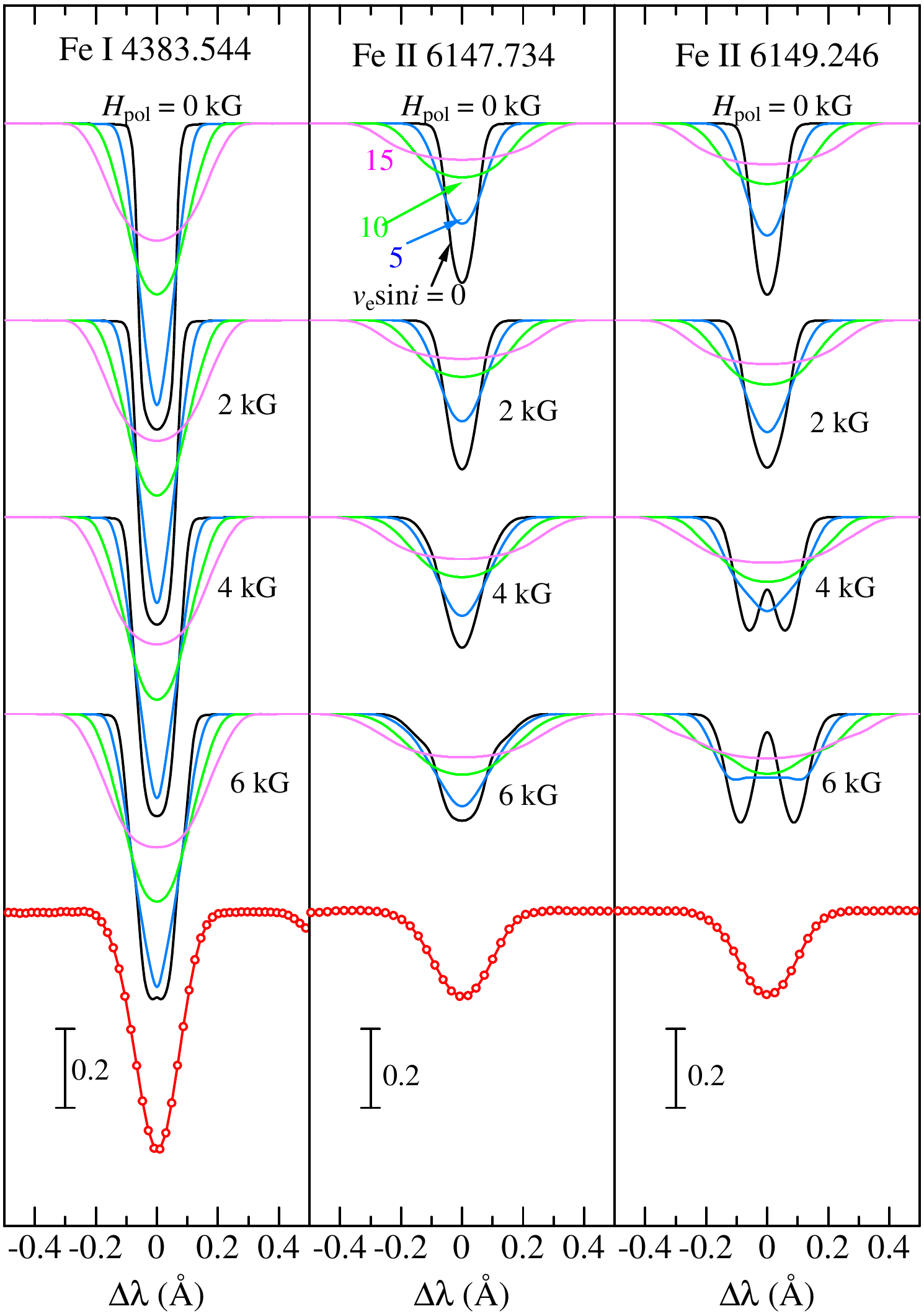}
\end{center}
\caption{
Demonstrative examples of theoretical flux profiles simulated by disk integration 
of unpolarized specific intensities (Stokes $I$) for three representative lines: 
Fe~{\sc i} 4383.544 (left), Fe~{\sc ii} 6147.734 (center), and Fe~{\sc ii} 6149.246 (right).
Shown here are results calculated with $v_{\rm t} = 1.5$~km~s$^{-1}$ 
for four $H_{\rm pol}$ values (0, 2, 4, and 6~kG) and four $v_{\rm e}\sin i$ values 
(0, 5, 10, and 15~km~s$^{-1}$). In addition, the actually observed profiles of 
$o$~Peg are also displayed at the bottom for comparison.
These simulated profiles of Fe~{\sc ii} 6147.734 and 6149.246 may be compared
with Fig.~2a and Fig.~2b of T91b, where the Zeeman-split structures are more manifest 
(because they are specific intensity profiles for single-valued magnetic field 
along with the assumption of $v_{\rm t} = 0$~km~s$^{-1}$ without rotational 
broadening).
}
\label{fig:6}
\end{minipage}
\end{figure*}

% Table 2
\setcounter{table}{1}
\begin{table*}[h]
\begin{minipage}{180mm}
\caption{Calculated $\sigma_{W}$ values as functions of $H_{\rm pol}$ and $v_{\rm t}$.}
%\small
\scriptsize
\begin{center}
\begin{tabular}{cccccccccc} 
\hline\hline
  $v_{\rm t}$  & $H_{\rm pol} = 0$ &   500 &  1000 &  1500 &  2000 &  2500 &  3000 &  3500 &  4000 \\
               &  $\langle H \rangle = 0$ &   321 &   642 &   962 &  1283 &  1604 &  1925 &  2246 &  2566 \\
\hline
  &  \multicolumn{9}{c}{(Fe~{\sc i} lines)} \\
 3.0  &  9.004 &  9.037 &  9.128 &  9.274 &  9.471 &  9.711 &  9.985 & 10.285 & 10.607 \\
 2.5  &  6.607 &  6.647 &  6.757 &  6.932 &  7.165 &  7.444 &  7.760 &  8.101 &  8.462 \\
 2.0  &  4.445 &  4.491 &  4.614 &  4.810 &  5.068 &  5.376 &  5.721 &  6.091 &  6.476 \\
 1.5  & \underline{3.164} &  3.192 &  3.273 &  3.414 &  3.615 &  3.873 &  4.176 &  4.511 &  4.866 \\
 1.0  &  3.552 &  3.530 &  3.483 &  3.438 &  3.428 &  3.473 &  3.581 &  3.748 &  3.961 \\
 0.5  &  4.613 &  4.557 &  4.420 &  4.239 &  4.057 &  3.909 &  3.815 &  3.786 &  3.823 \\
 0.0  &  5.105 &  5.037 &  4.870 &  4.647 &  4.411 &  4.199 &  4.036 &  3.935 &  3.901 \\
\hline
  &  \multicolumn{9}{c}{(Fe~{\sc ii} lines)} \\
 3.0  & 13.143 & 13.239 & 13.500 & 13.914 & 14.463 & 15.129 & 15.892 & 16.734 & 17.642 \\
 2.5  &  8.508 &  8.610 &  8.890 &  9.337 &  9.933 & 10.658 & 11.486 & 12.400 & 13.379 \\
 2.0  &  5.097 &  5.163 &  5.357 &  5.707 &  6.226 &  6.903 &  7.714 &  8.631 &  9.629 \\
 1.5  &  5.323 &  5.247 &  5.085 &  4.934 & \underline{4.903} &  5.078 &  5.491 &  6.122 &  6.925 \\
 1.0  &  8.094 &  7.933 &  7.543 &  7.021 &  6.477 &  6.017 &  5.740 &  5.717 &  5.973 \\
 0.5  & 10.537 & 10.334 &  9.838 &  9.158 &  8.403 &  7.669 &  7.041 &  6.595 &  6.398 \\
 0.0  & 11.481 & 11.261 & 10.730 & 10.001 &  9.186 &  8.376 &  7.655 &  7.095 &  6.759 \\
\hline
\end{tabular}
\end{center}
Given in this table are the values of $\sigma_{W}$ [standard deviation between the observed
and calculated equivalent widths in unit of m\AA; defined by Eq.~(1)] 
calculated for each combination of $H_{\rm pol}$ (field strength at the magnetic pole in unit of G; 
see the top row) and $v_{\rm t}$ (microturbulence in unit of km~s$^{-1}$; see the leftmost column). 
At the second row, the mean field strengths (in G) averaged over the stellar disk [$\langle H \rangle$; cf. Eq.(3)] corresponding to each $H_{\rm pol}$ are given. 
The minimum $\sigma_{W}$ among each group is indicated by an underline.
\end{minipage}
\end{table*}

% Table 3
\setcounter{table}{2}
\begin{table*}[h]
\begin{minipage}{180mm}
\caption{Calculated $\sigma_{h}$ values as functions of $H_{\rm pol}$ and $v_{\rm e}\sin i$.}
%\small
\scriptsize
\begin{center}
\begin{tabular}{cccccccccc} 
\hline\hline
  $v_{\rm e}\sin i$  & $H_{\rm pol} = 0$ &   500 &  1000 &  1500 &  2000 &  2500 &  3000 &  3500 &  4000 \\
               & $\langle H \rangle = 0$ &   321 &   642 &   962 &  1283 &  1604 &  1925 &  2246 &  2566 \\
\hline
  &  \multicolumn{9}{c}{(Fe~{\sc i} lines)} \\
     15.0  & 14.936 & 14.946 & 14.977 & 15.029 & 15.100 & 15.191 & 15.303 & 15.438 & 15.597 \\
     12.5  & 10.608 & 10.621 & 10.661 & 10.727 & 10.819 & 10.940 & 11.089 & 11.268 & 11.480 \\
     10.0  &  6.343 &  6.361 &  6.414 &  6.503 &  6.628 &  6.792 &  6.997 &  7.243 &  7.532 \\
      7.5  &  2.265 &  2.288 &  2.359 &  2.480 &  2.654 &  2.886 &  3.178 &  3.530 &  3.942 \\
      5.0  &  2.116 &  2.082 &  1.988 &  1.841 &  1.661 &  1.491 & \underline{1.405} &  1.485 &  1.762 \\
      2.5  &  5.133 &  5.079 &  4.928 &  4.677 &  4.337 &  3.929 &  3.490 &  3.066 &  2.724 \\
      0.0  &  6.127 &  6.073 &  5.916 &  5.645 &  5.263 &  4.793 &  4.274 &  3.751 &  3.280 \\
\hline
  &  \multicolumn{9}{c}{(Fe~{\sc ii} lines)} \\
     15.0  & 13.736 & 13.749 & 13.789 & 13.853 & 13.940 & 14.050 & 14.184 & 14.344 & 14.532 \\
     12.5  &  9.675 &  9.691 &  9.739 &  9.818 &  9.928 & 10.069 & 10.244 & 10.454 & 10.702 \\
     10.0  &  5.684 &  5.705 &  5.767 &  5.871 &  6.017 &  6.207 &  6.445 &  6.732 &  7.071 \\
      7.5  &  1.937 &  1.963 &  2.040 &  2.175 &  2.372 &  2.638 &  2.977 &  3.389 &  3.873 \\
      5.0  &  2.149 &  2.109 &  1.999 &  1.829 &  1.631 &  1.466 & \underline{1.442} &  1.645 &  2.063 \\
      2.5  &  4.736 &  4.675 &  4.503 &  4.221 &  3.847 &  3.421 &  3.005 &  2.684 &  2.548 \\
      0.0  &  5.588 &  5.525 &  5.344 &  5.034 &  4.611 &  4.120 &  3.629 &  3.207 &  2.923 \\
\hline
\end{tabular}
\end{center}
Given in this table are the values of $\sigma_{h}$ [standard deviation between the observed
and calculated full-width at half-maximum in unit of km~s$^{-1}$; defined by Eq.~(4)] 
calculated for each combination of $H_{\rm pol}$ (field strength at the magnetic pole in unit of G; 
see the top row) and $v_{\rm e}\sin i$ (projected rotational velocity in unit of km~s$^{-1}$; see the leftmost column). 
The minimum $\sigma_{h}$ among each section is indicated by an underline.
Otherwise, the same as in Table~2.
\end{minipage}
\end{table*}

%Sect. 5
\section{Magnetic field determination}

\subsection{Equivalent widths analysis}

Let us first try to establish ($H_{\rm pol}$, $v_{\rm t}$) from equivalent widths ($W$). 
Here, Method~2 described in Sect.~3.3 is applied, in which $W_{\rm cal}^{7.8}$ 
(theoretical equivalent width calculated with $A = 7.8$)\footnote{
The integrated strengths (equivalent widths) of unsaturated weak lines in the linear 
part of the curve of growth, which essentially determine the abundance, are practically 
free from any Zeeman broadening effect (like the effect of microturbulence). 
Accordingly, the Fe abundance of $A = 7.8$ derived in Sect.~3.2 by the conventional 
analysis is invariably valid irrespective of the existence of any magnetic field.} 
is compared with $W_{\rm obs}$. 
Since theoretical $W_{\rm cal}^{7.8}$ data are prepared for combinations of 
$H_{\rm pol}$ and $v_{\rm t}$ (while results for $v_{\rm e}\sin i = 0$ were 
adopted because of its irrelevance in this case), $\sigma_{W}$ defined by Eq.~(1) 
is also regarded as a function of these two parameters.

The resulting $\sigma_{W}(H_{\rm pol}, v_{\rm t})$ values derived from 
Fe~{\sc i} and Fe~{\sc ii} lines are given in Table~2, and the contours of 
$\sigma_{W}$ on the $H_{\rm pol}$--$v_{\rm t}$ plane are depicted in Fig.~7.
As seen from the locations of minimum $\sigma_{W}$, $H_{\rm pol}$ solutions 
for Fe~{\sc i} ($\sim 0$~G) and Fe~{\sc ii} ($\sim 2000$~G) are 
rather conflicting, though $v_{\rm t} \sim 1.5$~km~s$^{-1}$ is consistently
obtained irrespective of the species. 

%Fig. 7
\setcounter{figure}{6}
\begin{figure}[h]
\begin{minipage}{85mm}
\begin{center}
  \includegraphics[width=8.5cm]{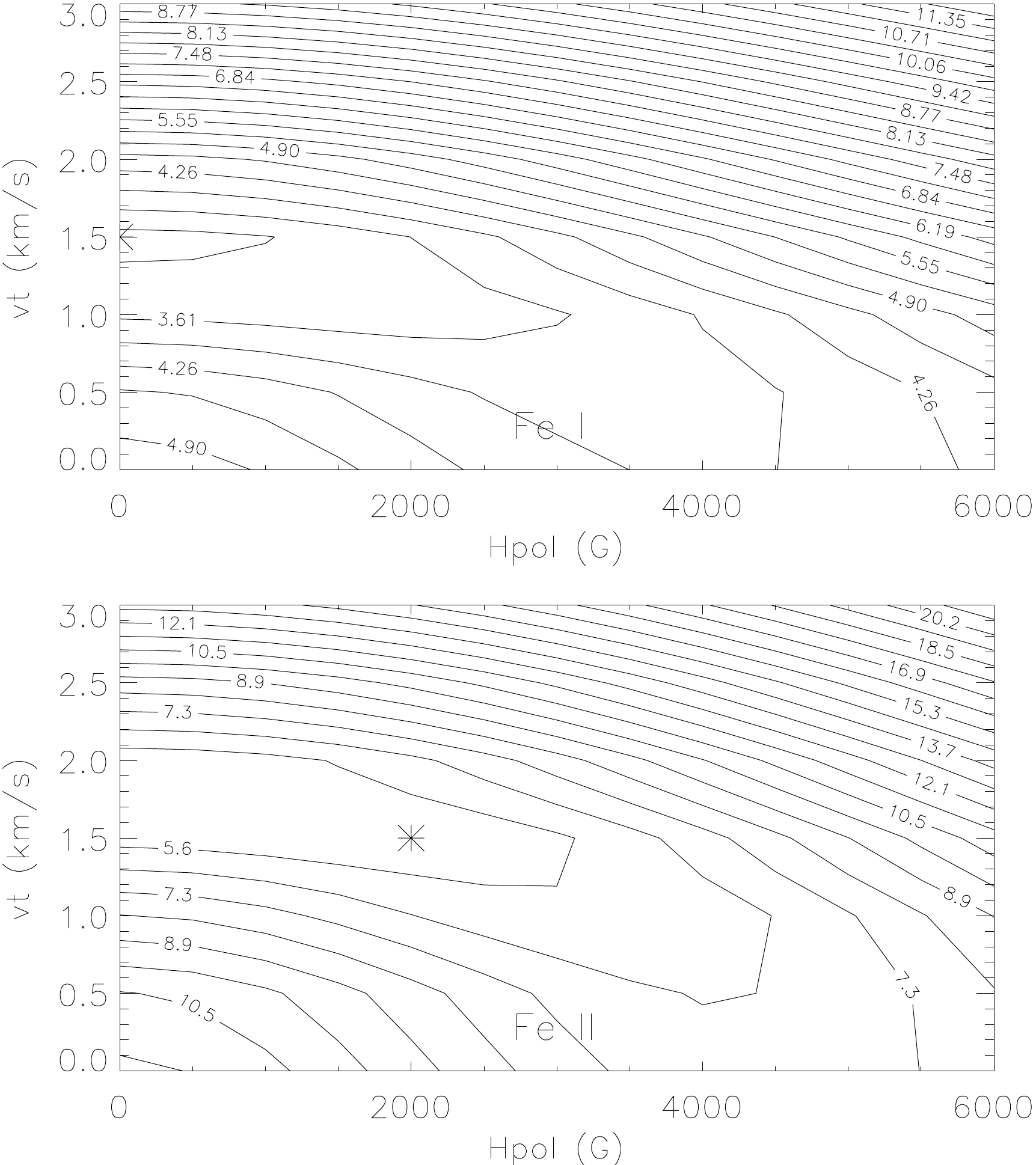}
\end{center}
\caption{
Graphical display of the contours of $\sigma_{W}$ on the
$H_{\rm pol}$--$v_{\rm t}$ plane,
where the results for Fe~{\sc i} and Fe~{\sc ii} lines are separately 
displayed in the upper and lower panels, respectively.
In each panel, the position of ($H_{\rm pol}^{*}$, $v_{\rm t}^{*}$) 
corresponding to the minimum $\sigma_{W}$ is indicated by an asterisk (*).
}
\label{fig:7}
\end{minipage}
\end{figure}

\subsection{Line widths analysis}

Next, we extract information of magnetic field from the line width 
($h$), where the contribution of $v_{\rm e}\sin i$ plays an important role.
For this purpose, the observed width ($h_{\rm obs}$) is compared with the 
theoretical width ($h_{\rm cal}^{7.8}$) calculated for various combinations 
of $H_{\rm pol}$ and $v_{\rm e}\sin i$ but for fixed $A = 7.8$ and 
$v_{\rm t} = 1.5$~km~s$^{-1}$ (according to the result of Sect.~5.1).
Similarly to Eq.~(1), we define $\sigma_{h}$ (function of $H_{\rm pol}$ 
and $v_{\rm e}\sin i$) as
\begin{equation}
\sigma_{h} \equiv 
\sqrt{\sum_{n=1}^{N}(h_{v,{\rm cal},n}^{7.8} - h^{0}_{v,{\rm obs},n})^{2}/N}.
\end{equation}
Here, $h^{0}_{v,{\rm obs}} \equiv \sqrt {h_{v,{\rm obs}}^{2}- 3^{2}}$
is the observed line width (in km~s$^{-1}$) corrected for the instrumental effect
(FWHM of 3~km~s$^{-1}$), where the fact that line profiles are well 
approximated by Gaussian function (cf. Fig.~2e) was taken into account.

The resulting $\sigma_{h}(H_{\rm pol}, v_{\rm e}\sin i)$ values derived from 
Fe~{\sc i} and Fe~{\sc ii} lines are given in Table~3, and the contours of 
$\sigma_{h}$ on the $H_{\rm pol}$--$v_{\rm e}\sin i$ plane are depicted in Fig.~8.
Inspecting the locations of minimum $\sigma_{h}$, we obtain  
$H_{\rm pol} \sim 3000$~G and $v_{\rm e}\sin i \sim 5$~km~s$^{-1}$ 
for both Fe~{\sc i} and Fe~{\sc ii} lines.  

%Fig. 8
\setcounter{figure}{7}
\begin{figure}[h]
\begin{minipage}{85mm}
\begin{center}
  \includegraphics[width=8.5cm]{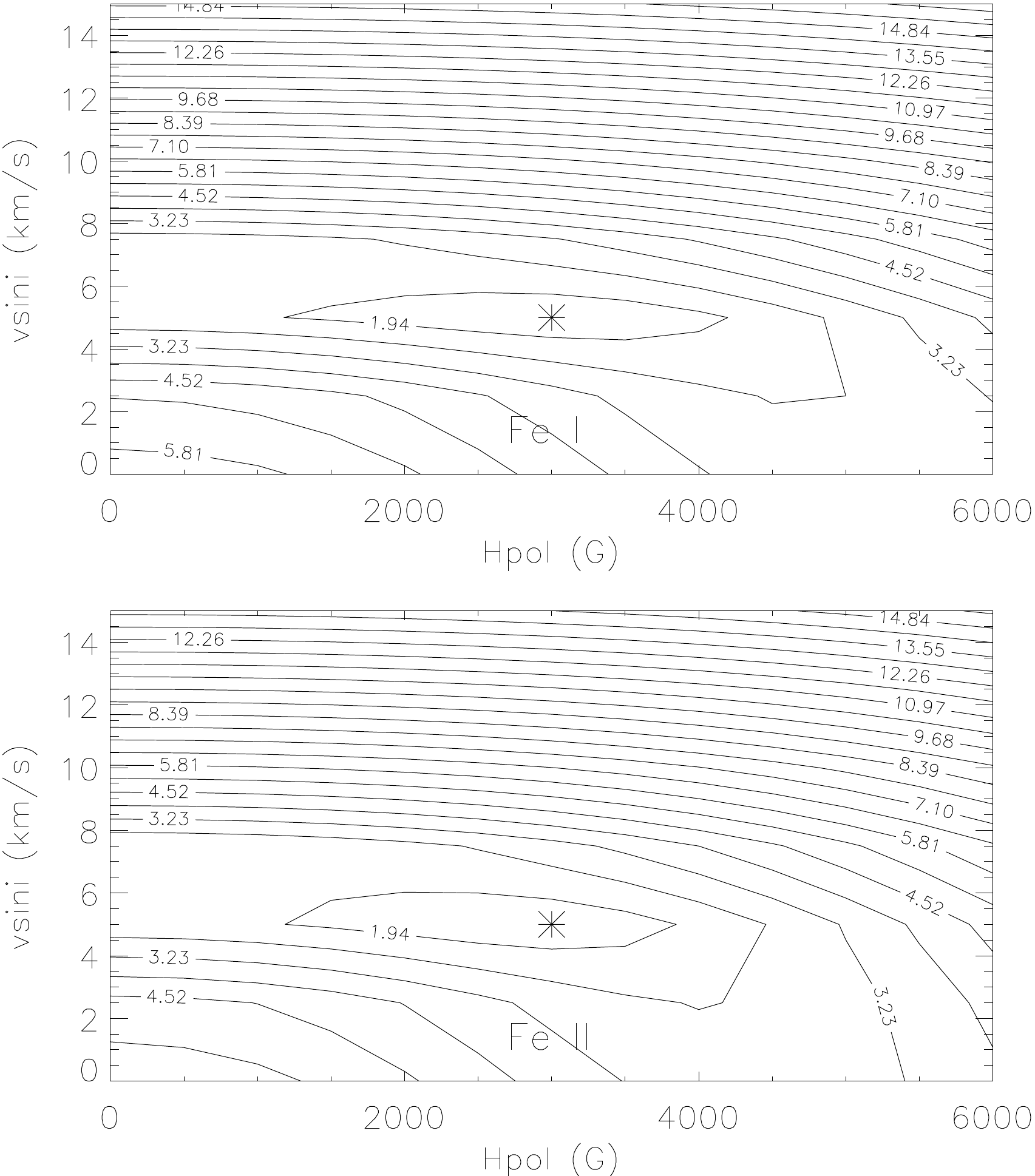}
\end{center}
\caption{
Graphical display of the contours of $\sigma_{h}$ on the
$H_{\rm pol}$--$v_{\rm e}\sin i$ plane, 
where the results for Fe~{\sc i} and Fe~{\sc ii} lines are separately 
displayed in the upper and lower panels, respectively.
In each panel, the position of ($H_{\rm pol}^{*}$, $v_{\rm e}\sin i^{*}$) 
corresponding to the minimum $\sigma_{h}$ is indicated by an asterisk (*).
}
\label{fig:8}
\end{minipage}
\end{figure}

% Table 4
\setcounter{table}{3}
\begin{table}[h]
\begin{minipage}{85mm}
\caption{Summary of solutions based on line-strength or line-width analysis.}
%\small
\begin{center}
\begin{tabular}{ccccc} 
\hline\hline
\multicolumn{5}{c}{(Line strengths analysis)}\\
Species & $H_{\rm pol}^{*}$ & $\langle H \rangle^{*}$ & $v_{\rm t}^{*}$ & $v_{\rm t}^{\rm est}$ \\
\hline
 Fe~{\sc i} &  0  &  0  & 1.5  & 1.37 \\
 Fe~{\sc ii}& 2000 & 1283 & 1.5 & 1.52 \\
\hline\hline
\multicolumn{5}{c}{(Line widths analysis)}\\
Species & $H_{\rm pol}^{*}$ & $\langle H \rangle^{*}$ & $v_{\rm e}\sin i^{*}$ & $v_{\rm e}\sin i^{\rm est}$ \\
\hline
 Fe~{\sc i}  &  3000  &  1925  & 5.0  & 5.10 \\
 Fe~{\sc ii} & 3000 & 1925 & 5.0 & 5.01 \\
\hline
\end{tabular}
\end{center}
Quantities with asterisks (*) in column 2--4 are the solutions corresponding to the minimum of 
$\sigma_{W}$ or $\sigma_{h}$, 
while that in column~5 is the estimated solution derived by quadratic interpolation (see Sect.~6.1).
\end{minipage}
\end{table}

%Sect. 6
\section{Discussion}

\subsection{Results and their characteristics}

In Sect.~5, we derived the magnetic field of $o$~Peg ($H_{\rm pol}$ or $\langle H \rangle$) 
and the related line-broadening parameters ($v_{\rm t}$ and $v_{\rm e}\sin i$) 
by comparing the observed and simulated equivalent widths ($W$) and line widths ($h$).
The results are summarized in Table~4.

Since the velocity parameter solutions ($\eta^{k^{*}}$ at the grid node 
$k = k^{*}$, where $\eta$ denotes either $v_{\rm t}$ or $v_{\rm e}\sin i$) 
have rounded values because the grids are rather coarse, 
$\sigma$ was analytically expressed by a second-order polynomial of $\eta$ by 
using $\sigma(k^{*}-1)$, $\sigma(k^{*})$, and $\sigma(k^{*}+1)$, from which 
the new $\eta$ solution ($\eta^{\rm est}$) was estimated as 
corresponding to the minimum of this parabolic $\sigma(\eta)$.
Such derived $v_{\rm t}^{\rm est}$ and $v_{\rm e}\sin i^{\rm est}$
are also given in Table~4.

By inspecting these tables, we can read the following consequences regarding  
the magnetic field strength $\langle H \rangle$ (as well as $v_{\rm t}$ and 
$v_{\rm e}\sin i$) of $o$~Peg.
\begin{itemize}
\item
Regarding the equivalent width analysis, contradicting results 
are obtained for $\langle H \rangle$ ($\sim 0$~kG from Fe~{\sc i} lines 
and $\sim 1.3$~kG from Fe~{\sc ii} lines). However, since the former 
is likely to be less reliable for the reason described in Sect.~3.3, we 
preferentially adopt the latter solution of $\langle H \rangle \sim 1.3$~kG. 
Meanwhile, $v_{\rm t}$ is consistently settled at $\sim 1.5$~km~s$^{-1}$.   
\item
As to the line width analysis, mean field strengths of 
$\langle H \rangle \sim 1.9$~kG are derived for both Fe~{\sc i} and 
Fe~{\sc ii} lines. The projected rotational velocity is concluded
to be $v_{\rm e}\sin i \sim 5$~km~s$^{-1}$.
\item
Based on these results, although the $\langle H \rangle$ value from $W$ 
tends to be somewhat lower than that from $h$, the mean magnetic field on 
the order of $\langle H \rangle \sim$~1.5--2~kG in $o$~Peg is anyhow confirmed. 
Accordingly, the consequence of our new analysis is almost consistent with 
the conclusion of previous studies (ML90, T91b, T93), which reported the 
existence of $H \sim$~2~kG in this star. 
\item
We may state that the impact of magnetic field is not very significant on 
the spectroscopic determination of $v_{\rm t}$ and $v_{\rm e}\sin i$, because 
the resulting values ($\sim 1.5$~km~s$^{-1}$ and $\sim 5$~km~s$^{-1}$) 
are not much different from those derived by neglecting the magnetic effect
($v_{\rm t}\sim$~1.7--1.8~km~s$^{-1}$ derived in Sect.~3.2, and 
the typical recent literature values of $v_{\rm e}\sin i$ are 
$\sim$~6--7~km~s$^{-1}$ as seen in Table~1). 
Regarding $v_{\rm t}$, this is a reconfirmation of the argument in T93 
(but not that in T91b).  
\end{itemize}

\subsection{Precision check of T93 approximation}

In the analysis of equivalent widths (Sect.~5.1), we could establish the 
magnetic field of $o$~Peg from Fe~{\sc ii} lines ($H_{\rm pol} \sim$~1.5--2~kG), 
but not from Fe~{\sc i} lines (i.e., a well-defined minimum is not found 
in $\sigma_{W}$ which continues to decline with a decrease in $H_{\rm pol}$ 
until $H_{\rm pol} \rightarrow 0$). 
This situation is rather similar to the case of T93, where a successful
result was obtained from Fe~{\sc ii} lines but not from Fe~{\sc i} lines
(see the run of $\sigma_{\rm b}$ depicted in the middle-row panels
of Fig.~2 in T93). 

In T93, a practical (but approximate) method was used for evaluating the 
line flux equivalent width under the existence of magnetic field, in which 
the conventional spectrum-synthesis code is applicable without any necessity 
of solving the transfer equation of polarized radiation (cf. Sect.~2 in T93 
for a detailed explanation). Briefly speaking, in this method, two equivalent 
widths are calculated for a given $H$ corresponding to the minimum intensification 
($W_{\rm a}$; using only $\sigma_{-}$ and $\sigma_{+}$ components but 
independently from each other) and maximum intensification ($W_{\rm c}$; 
use of $\sigma_{-}$, $\sigma_{+}$, and $\pi$ components altogether while 
assuming as if no polarization effect exists). Then, it was assumed in T93 
that the theoretical equivalent width to be adopted (which should be between 
$W_{\rm a}$ and $W_{\rm c}$) is given by the ``simple mean'' of these 
minimum and maximum as $W_{\rm b}(H) \equiv [W_{\rm a}(H) + W_{\rm c}(H)]/2$.  

In order to examine the precision of this approximation, $W_{\rm b}$ values 
were calculated at various field strengths ($H$) for all of the 380 Fe lines 
(with $A= 7.8$ and $v_{\rm t} = 1.5$~km~s$^{-1}$), which were then compared 
with the corresponding $W_{\rm cal}$ values simulated in Sect.~4.3.

The resulting $W$ vs. $H$ relations (based on different methods of 
T93 and this study) for three representative lines (Fe~{\sc i} 4383.544, 
Fe~{\sc ii} 6147.734, and Fe~{\sc ii} 6149.246) are compared in 
Figs.~9a, 9b, and 9c, respectively.
We can see from these figures that $W_{\rm b}(H)$ (solid line) is a 
reasonable approximation of $W_{\rm cal}(\langle H \rangle)$ (symbols), though 
some systematic departure is observed at larger $H$ depending on lines.\footnote{
We may state that a line with single (or practically single) $\sigma_{-}$ 
or $\sigma_{+}$ component (such like the cases of Fe~{\sc i} 4383.544 or 
Fe~{\sc ii} 6149.246) tends to suffer an appreciable deviation, since the 
difference between $W_{\rm a}$ and $W_{\rm c}$ is comparatively large 
because $W_{\rm a}$ is $H$-independent and constant (cf. Figs.~9a and 9c).
In contrast, if $\sigma_{-}$ (or $\sigma_{+}$) components of a line are 
sufficiently apart (like Fe~{\sc ii} 6147.734), $W_{\rm b}$ makes a fairly good 
approximation for $W_{\rm cal}$, because $W_{\rm a}$ and $W_{\rm c}$ increase with 
$H$ in somewhat similar manner and the difference tends to be small (cf. Fig.~9b).}  
The logarithmic differences between $W_{\rm b}$ and $W_{\rm cal}$ for all lines
are plotted against $W_{\rm cal}$ in Figs.~9d, 9e, and 9f for different 
field strengths of 0, 2, and 4~kG, respectively.  
These figures indicate that $|\log (W_{\rm b}/W_{\rm cal})|$ is typically 
a few hundredths dex at most (i.e., several or $\lesssim 10$ percent in $W$) 
even at the magnetic field of 4~kG.
Accordingly, the practical approach proposed by T93 for calculating equivalent 
widths of a magnetic star may be regarded as a reasonable approximation of
moderate precision (especially when the lines to be used are carefully chosen). 

%Fig. 9
\setcounter{figure}{8}
\begin{figure*}[h]
\begin{minipage}{180mm}
\begin{center}
  \includegraphics[width=12cm]{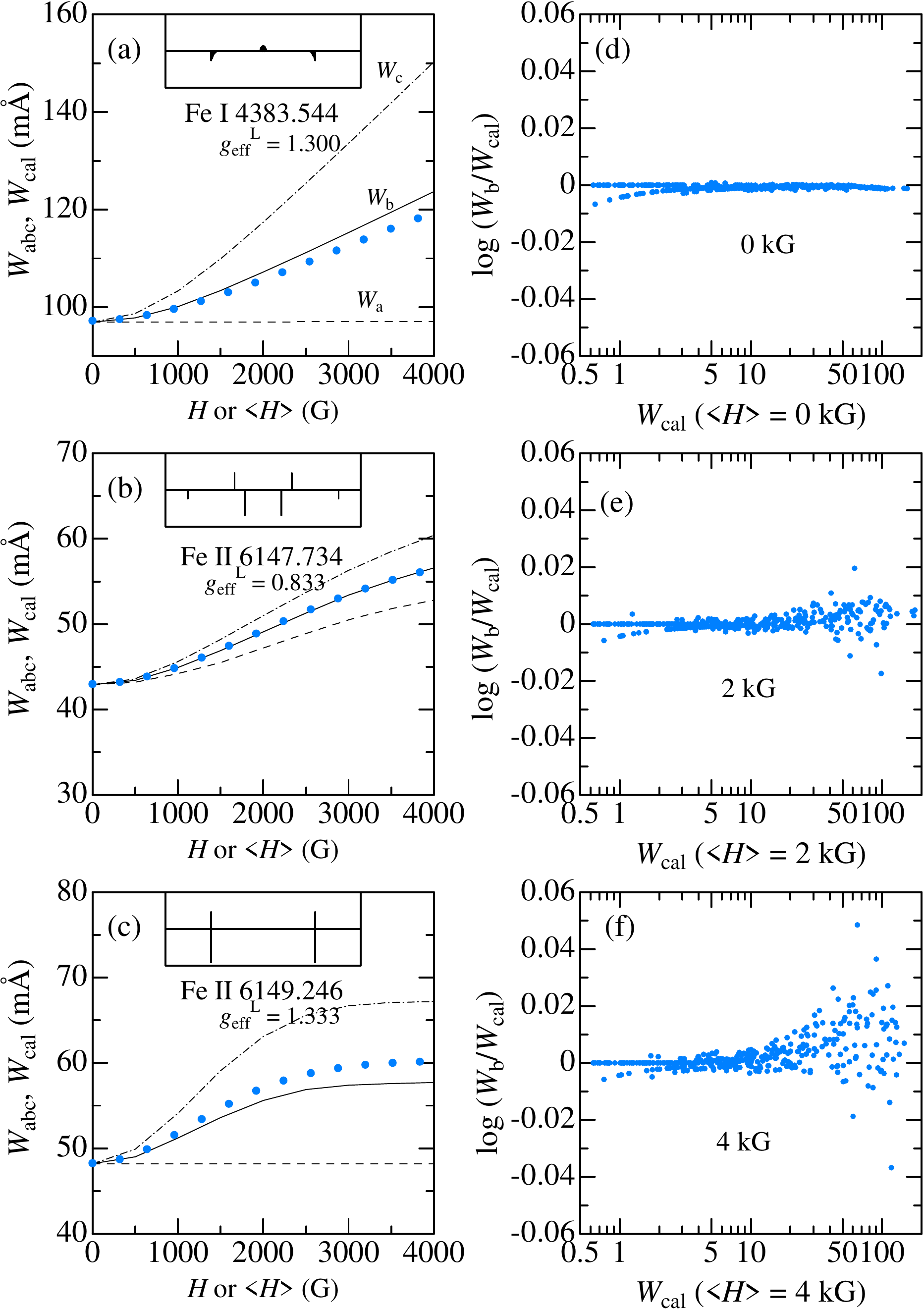}
\end{center}
\caption{
Left panels (a--c) show how the flux equivalent width varies by changing
the magnetic field strength for three representative lines:
(a) Fe~{\sc i} 4383.544, (b) Fe~{\sc ii} 6147.734, and (c) Fe~{\sc ii} 6149.246
(their Zeeman patterns are shown in the inset of each panel). 
Those evaluated by the WIDTH9 program with three kinds of approximations proposed 
in T93, $W_{\rm a}$ (minimum intensification involving only $\sigma_{-}$ and 
$\sigma_{+}$ components), $W_{\rm c}$ (maximum intensification for the case of 
neglecting the polarization effect), and $W_{\rm b}$ (simple mean of $W_{\rm a}$ 
and $W_{\rm c}$; finally adopted in T93), are depicted in dashed line, dash-dotted line, 
and solid line, respectively. Meanwhile, those calculated based on our dipole magnetic 
field model by disk integration of local $I$ profiles ($W_{\rm cal}$) are plotted 
by filled symbols.
Note that these $W_{\rm cal}$ values are plotted against 
$\langle H \rangle$ (not $H_{\rm pol}$). All these $W$ calculations 
were done with $A = 7.80$ and $v_{\rm t} = 1.5$~km~s$^{-1}$.
The logarithmic differences evaluated for all 380 Fe lines
[$\log (W_{\rm b}/W_{\rm cal})$] are plotted against 
$W_{\rm cal}$ in the right panels (d--f) for different mean field strengths
($\langle H \rangle$): (d) 0~kG, (e) 2~kG, and (f) 4~kG. 
}
\label{fig:9}
\end{minipage}
\end{figure*}

%Fig. 10
\setcounter{figure}{9}
\begin{figure}[h]
\begin{minipage}{70mm}
\begin{center}
  \includegraphics[width=7cm]{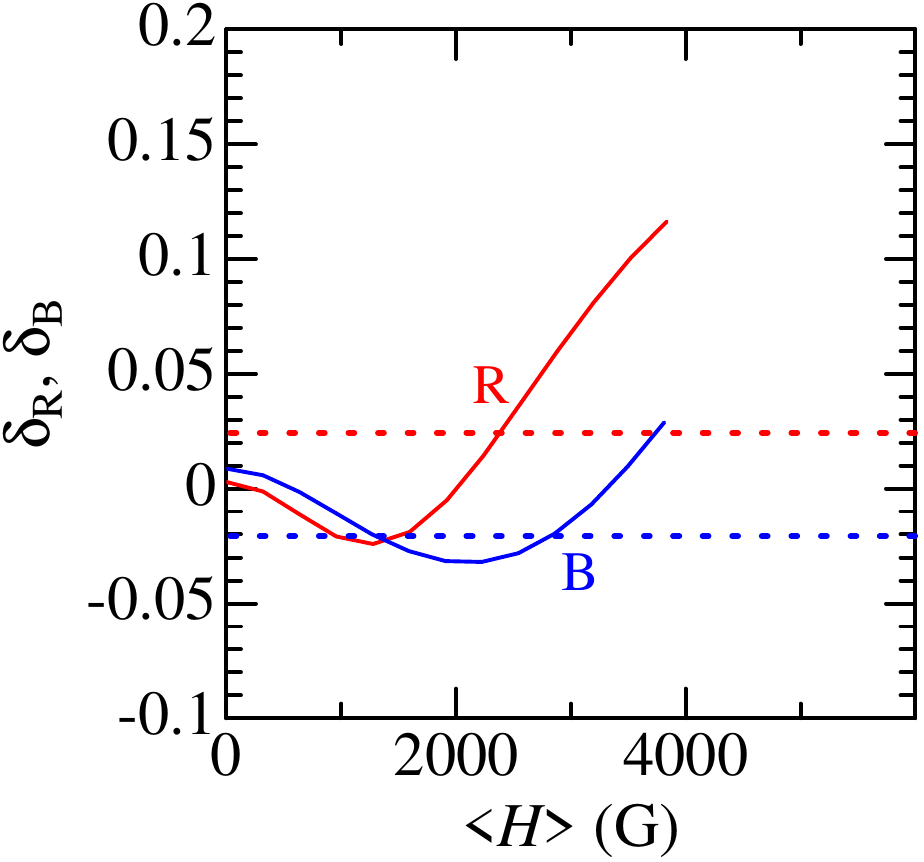}
\end{center}
\caption{
Relative differences of equivalent widths [$\delta \equiv 2(W_{1}-W_{2})/(W_{1}+W_{2})$]
for the red (R) and blue (B) pair lines (cf. Table~1 in T91b) are plotted against 
the mean field strength ($\langle H \rangle$) by solid lines, which were calculated 
with $A = 7.80$ and $v_{\rm t} = 1.5$~km~s$^{-1}$. 
The observed values ($\delta_{\rm R} = +0.025$, $\delta_{\rm B} = -0.020$)
are indicated by horizontal dotted lines.
}
\label{fig:10}
\end{minipage}
\end{figure}

\subsection{Implication from the line-pair method}

Finally, as an application of the simulations done in Sect.~4, we estimate 
the magnetic field of $o$~Peg based on the very simple approach using the 
strengths of specific line pairs belonging to the same multiplet, 
which was first tried by ML90 and then extended by T91b.
This method makes use of the relative difference of equivalent widths for 
the two lines (1 and 2) defined as $\delta \equiv 2(W_{1}-W_{2})/(W_{1}+W_{2})$. 
While $\delta \simeq 0$ in the non-magnetic case, $\delta$ begins to depart 
from zero with an increase in $H$ (because of the different $H$-sensitivity 
between lines 1 and 2). Accordingly, $H$ may be estimated by comparing the 
observed $\delta_{\rm obs}$ with the known $\delta$ vs. $H$ relation.
 
Here, two line pairs are relevant, which are called after T91b as ``red pair (R)''
(Fe~{\sc ii} 6147.7 and 6149.2) and ``blue pair (B)'' (Fe~{\sc ii} 4416.8 and 4385.4).
See Table~1 of T91b for more details on these line pairs.
Since these 4 Fe~{\sc ii} lines are included in our 380 target lines,   
$\delta_{\rm R}$ and $\delta_{\rm B}$ at various field strengths can be 
evaluated from their $W_{\rm cal}$ results\footnote{ 
Since the $\log gf$ values of the lines consisting the pair given 
in the VALD database (which we adopted in this study) are rather discrepant
from each other, the requirement of $\delta \simeq 0$ in the non-magnetic case
is not fulfilled if $\log gf$(VALD) data were used. Therefore, the $\log gf$ 
values presented in Table~1 of T91b were exceptionally employed here
for both of the red-pair lines and blue-pair lines.}
calculated at $H_{\rm pol}$ = 0, 500, 1000, $\ldots$, 5500, and 6000~G 
(along with $v_{\rm t} = 1.5$~km~s$^{-1}$ and $A = 7.8$).

The resulting theoretical $\delta_{\rm R}$ and $\delta_{\rm B}$ 
are plotted against $\langle H \rangle$ in Fig.~10, where the positions of 
$\delta_{\rm R} = +0.025$ and $\delta_{\rm B} = -0.020$ derived from the observed 
equivalent widths in m\AA\ ($W_{1,{\rm R}}/W_{2,{\rm R}} = 49.8/48.6$, 
$W_{1,{\rm B}}/W_{2,{\rm B}} = 91.5/93.4$) are also indicated.
The following trends can be read from this figure.\\
--- A comparison of theoretical and observed $\delta_{\rm R}$ yields
a mean magnetic field strength of $\langle H \rangle \sim 2.4$~kG.\\
--- Regarding $\delta_{\rm B}$, a unique solution can not be found. 
What can be said from Fig.~10 is that $\langle H \rangle$ is either
$\sim 1.3$~kG or $\sim 2.9$~kG.\\
--- In any case, these results do not contradict the consequence of the 
main analysis (detection of $\langle H \rangle$ on the order of $\sim 2$~kG;
cf. Sect.~6.1).

\subsection{Magnetic nature of $o$~Peg}

Our analysis on the line strengths and widths has 
thus corroborated that an appreciable magnetic field of $\langle H \rangle \sim $1--2~kG 
(mean field strength averaged over the disk) exists in the Am star $o$~Peg.
Here, we should recall that previous spectropolarimetric observations 
conducted so far failed to detect any meaningful signal of circular polarization 
in this star (cf. Sect.~1), which means that $\langle H_{z}\rangle$ 
(mean line-of-sight component of the field averaged over the disk) is very weak. 
Although detection of ultra-weak $\langle H_{z} \rangle$ on the order of 
several G might as well be possible by using higher-precision observations 
(see footnote~1), we can at least state 
that $\langle H_{z} \rangle$ is negligibly weak compared to $\langle H \rangle$.

One possibility to explain this marked disagreement is that the magnetic field is not 
globally organized but has a complex structure (such as suggested by ML90).
If several or more strong magnetic regions of smaller scale with different polarities 
exist on the stellar disk (such as sunspots), the net line-of-sight component
of the field ($\langle H_{z} \rangle$) would almost vanish while the mean magnetic field 
strength ($\langle H \rangle$) still remains detectable. However, it seems that very 
strong magnetic spots or patches (with strengths considerably exceeding $\sim 2$~kG) 
are rather unlikely in the present case, because they should give rise to some kind 
of appreciable peculiarities in the profiles of magnetically-sensitive lines.
For example, if assumed that 1/3 of the stellar disk is covered by a strong magnetic 
patch of $H \sim 6$~kG while the remaining 2/3 is non-magnetic, Fe~{\sc ii} 6149.246 
line would show a complex profile as expected from the simulation; but such an  
anomalous feature is absent in the actual profile which is nearly Gaussian (cf. 
the right panel of Fig.~6). 

Accordingly, whichever configuration of the magnetic field, the field contrast over 
the stellar disk would not be distinctly large (i.e., not so much like spots/patches 
as rather gradual). In this context, the simple rotating dipole model 
of aligned rotational/magnetic axis viewed almost equator-on 
($i \simeq \alpha \simeq 90^{\circ}$), which was assumed in the simulation of 
this study, may be regarded as the likely solution for $o$~Peg, because 
it naturally explains the observational fact of $\langle H_{z} \rangle \sim 0$. 
Although it is not easy to check this hypothesis observationally,
some weak rotational modulation of circular polarization might as well be detected 
if $\alpha$ is not exactly (but slightly deviates from) $90^{\circ}$. 
In this case, the rotation period is estimated as $P \simeq 42$~d by combining 
$v_{\rm e} (\simeq v_{\rm e}\sin i) \simeq 5$~km~s$^{-1}$ and $R \simeq 4.4 R_{\odot}$. 
It may thus be worthwhile to examine whether a modulation period of $\sim 40$~d 
is observed for this star by ultra high-precision spectropolarimetry. 

%Sect. 7
\section{Summary and conclusion}

The star $o$~Peg is a representative A-type star (classified as a hot Am star
from its abundance characteristics), which has been frequently studied by a number 
of investigators because of its brightness and sharp-line nature.

In the early 1990s, several authors (ML90, T91b, T93) reported the existence of 
surface magnetic field on the order of $\sim 2$~kG in this star based on the 
analysis of widths or strengths of many spectral lines, which was a significant finding 
because the conventional spectropolarimetry has been 
unsuccessful in detecting any meaningful signal of $\langle H_{z} \rangle$. 

However, the techniques employed by these old studies were not necessarily founded 
on a physically legitimate basis but rather empirical or approximate in character.
Besides, the quality of the adopted spectra of $o$~Peg, on which the observational 
data of line widths and strengths were measured, was not satisfactory as viewed
from the present-day standard.
 
Given that this detection does not seem to have been corroborated since then, 
I decided to revisit this issue based on (i) an improved modeling
of theoretical line flux profile of a rotating magnetic star 
(by disk integration of local intensity profiles obtained by correctly solving 
the transfer equation of polarized radiation) and (ii) using the high-resolution 
($R\sim 100000$) and very high S/N ($\sim$~1000) spectra of $o$~Peg.  

The magnetic and rotational axes of this model star (with a dipole field) 
were assumed to be in line with each other and perpendicular to the observer's 
line of sight ($i = \alpha = 90^{\circ}$), by which $\langle H_{z} \rangle = 0$
is attained in accordance with observations.

As for the spectral lines whose full-widths at half-maximum ($h$) and equivalent 
widths ($W$) are to be analyzed, 380 Fe lines (198 Fe~{\sc i} and 182 Fe~{\sc ii} 
lines) were carefully selected, which are free from any serious blending effect.  

The conventional analysis (without taking into account the effect of magnetic 
intensification) of equivalent widths was first carried out, which resulted in
$A \simeq 7.8$ (Fe abundance) and $v_{\rm t} \simeq$~1.7--1.8~km~s$^{-1}$ 
(microturbulence). This result of $A$(Fe) = 7.8 was used as the fiducial
abundance to be fixed throughout the subsequent magnetic field analyses. 

By requiring the minimum dispersion between the theoretical $W$ values 
simulated with the magnetic field model (depending on the field strength 
and microturbulence) and the observed ones, we found that 
$\langle H \rangle \sim 1.3$~kG (from Fe~{\sc ii} lines)
and $v_{\rm t} \sim 1.5$~km~s$^{-1}$.

Similarly, by comparing the simulated $h$ values (function of projected rotational
velocity and magnetic field strength) with the measured ones, the best solutions
accomplishing the least dispersion were derived as 
$\langle H \rangle \sim 1.9$~kG and $v_{\rm e}\sin i \sim 5$~km~s$^{-1}$.

Based on these results, although the $\langle H \rangle$ value from the
analysis of $W$ tends to be somewhat lower than that from $h$, 
the mean magnetic field on the order of $\langle H \rangle \sim$~1--2~kG 
has been confirmed in $o$~Peg. 

In addition, supplementary applications of the simulated $W$ results were 
also conducted for checking purposes: 
(i) The precision of the practical method proposed by T93 for evaluating $W$ 
in the presence of a magnetic field was examined and confirmed to be a 
reasonable and useful approximation. 
(ii) The line-pair method used by ML90 and T91b was applied
based on the newly simulated $W$ values of specific line pairs
and found that $H$ is in the range of $\sim$~1--3~kG.  

In summary, the consequence resulting from our analysis on the $W$ and 
$h$ data of Fe lines is almost consistent with the conclusion of previous 
studies (ML90, T91b, T93) which reported $H \sim$~2~kG for this star. 

Regarding the reason for the marked discrepancy between $\langle H \rangle \sim$~(1--2~kG) 
and $\langle H_{z} \rangle (\sim 0)$, 
an accidental accomplishment of $i \simeq \alpha \simeq 90^{\circ}$ in 
the poloidal configuration might as well be ponderable, rather than invoking 
a complex structure with small-scale magnetic regions of different polarities. 

\section*{Acknowledgments}

This research has made use of the SIMBAD database, operated by CDS, 
Strasbourg, France. 
This work has also made use of the VALD database, operated at Uppsala 
University, the Institute of Astronomy RAS in Moscow, and the University of Vienna.

\section*{Data availability}

The basic data and results underlying this article are presented as 
the online supplementary material. The line profile data used for 
measurements are given in ``obsprofiles.dat'', while the original 
spectra of $o$~Peg are in the public domain and available at 
https://smoka.nao.ac.jp/index.jsp (SMOKA Science Archive site).

\section*{Supporting information}

This article accompanies the following online materials.
\begin{itemize}
\item
{\bf ReadMe.txt} 
\item
{\bf felines.dat} 
\item
{\bf obsprofiles.dat} 
\end{itemize}

\end{document}